\documentclass[11pt,aps,prd,preprint,groupedaddress,tightenlines,nofootinbib]{revtex4}
\usepackage{amsmath,amssymb,amsbsy,color,axodraw}
\usepackage{psfrag}  

\newcommand{\pslash}{p\kern-1ex /}
\newcommand{\lslash}{l\kern-1ex /}
\newcommand{\sslash}{s\kern-1ex /}
\newcommand{\Dslash}{{\cal D}\kern-1.5ex /}
\newcommand{\bpsi}{\overline{\psi}}

\newcommand{\be}{\begin{equation}}
\newcommand{\ee}{\end{equation}}
\newcommand{\bea}{\begin{eqnarray}}
\newcommand{\eea}{\end{eqnarray}}

\newcommand{\la}{\langle}
\newcommand{\ra}{\rangle}
\newcommand{\kzbar}{\bar{K}^0}
\newcommand{\msbar}{\overline{\rm MS}}
\newcommand{\beal}{\begin{align}}
\newcommand{\enal}{\end{align}}
\newcommand{\nn}{\nonumber}

\begin{document}

\begin{flushright}
{\normalsize UTHEP-516}\\
{\normalsize UTCCP-P-19}\\
\end{flushright}

\title{
Perturbative renormalization factors for generic $\Delta s=2$ 
four-quark operators in domain-wall QCD with improved gauge action
}
\author{Yousuke Nakamura$^{1}$ and Yoshinobu Kuramashi$^{1,2}$}
\affiliation{
$^1$Graduate School of Pure and Applied Sciences,\\ University of Tsukuba, Tsukuba 305-8571, Japan \\
$^2$Center for Computational Physics,\\ University of Tsukuba, Tsukuba 305-8577, Japan
}

\date{\today}

%
\begin{abstract}
%

We calculate one-loop renormalization factors of generic 
$\Delta s=2$ four-quark operators for domain-wall QCD with
the plaquette gauge action and the Iwasaki gauge action.
The renormalization factors are presented in the modified 
minimal subtraction ($\msbar$)
scheme with the naive dimensional regularization.
As an important application we show how to construct
the renormalization factors 
for the operators contributing to $K^0-\kzbar$ mixing  
in the supersymmetric models with the use of our results.

\end{abstract}

\maketitle

\section{Introduction}
%
Experimental studies of $K^0-\kzbar$ mixing provide us an opportunity
to deduce the indirect CP violation parameter $\epsilon_K$.
In the Standard Model (SM), the low-energy effective Hamiltonian contains
the dimension-six four-quark operator
${\cal O}_{LL}={\bar s}\gamma_\mu(1-\gamma_5) d\cdot
{\bar s}\gamma_\mu(1-\gamma_5) d$, and its hadronic matrix element
$\la\kzbar\vert{\cal O}_{LL} \vert K^0 \ra$ is required 
to determine $ \epsilon_K$ from the experimental results
of $K^0-\kzbar$ mixing. 
On the other hand, the physics beyond the SM involves the
four-quark operators with more general chiral structures.
For example, the relevant operators in supersymmetric 
models are\cite{fcnc_susy}
\bea
{\cal O}_1 &=&
\bar{s}^a\gamma_\mu(1-\gamma_5)d^a\bar{s}^b\gamma_\mu(1-\gamma_5)d^b,
\label{eq:op_susy1}\\
{\cal O}_2 &=&
\bar{s}^a(1-\gamma_5)d^a\bar{s}^b(1-\gamma_5)d^b,
\label{eq:op_susy2}\\
{\cal O}_3 &=&
\bar{s}^a(1-\gamma_5)d^b\bar{s}^b(1-\gamma_5)d^a, 
\label{eq:op_susy3}\\
{\cal O}_4 &=&
\bar{s}^a(1-\gamma_5)d^a\bar{s}^b(1+\gamma_5)d^b,
\label{eq:op_susy4}\\
{\cal O}_5 &=&
\bar{s}^a(1-\gamma_5)d^b\bar{s}^b(1+\gamma_5)d^a,
\label{eq:op_susy5}
\eea
where $a$ and $b$ are color indices.

Lattice QCD should be an ideal tool to determine
the above matrix elements from the first principles.
In the past decades, much effort have been devoted 
to the calculation of $\la\kzbar\vert{\cal O}_{LL} \vert K^0\ra $
by employing various quark and 
gauge actions\cite{bk_ks,bk_lee,bk_w,bk_orsay,bk_blum,bk_cppacs,bk_rbc,bk_tm,bk_boston,bk_ol}.
On the contrary, the matrix elements for the operators of 
Eqs.(\ref{eq:op_susy1})$-$(\ref{eq:op_susy5}) are
less studied compared to the case of 
${\cal O}_{LL}$\cite{Allton99,lellouch05}.   
Our aim is at a detailed study of them with the use of the
domain-wall quark formulation in lattice QCD, which 
is expected to realize full chiral symmetry at finite lattice spacing 
up to exponential fall-off of the explicit chiral symmetry breaking 
contributions\cite{Shamir93,Shamir95} and
have been successfully applied to the calculation of  
$\la\kzbar\vert{\cal O}_{LL} \vert K^0\ra$\cite{bk_blum,bk_cppacs,bk_rbc}.

The matrix elements calculated on the lattice should be converted
to those defined in some continuum regularization scheme
[e.g., the modified minimal subtraction scheme ($\msbar$)].
This is achieved by the finite renormalization relating
the lattice composite operators to the continuum counterparts
defined in some reguarization scheme. 
In this paper we present the perturbative results 
of the renormalization factors
for the complete set of $\Delta s=2$ four-quark operators
consisting of physical quark fields in the domain-wall QCD (DWQCD).
With the use of our results one can obtain the renormalization factor
for an arbitrary $\Delta s=2$ four-quark operator.   
This work is an extention of Refs.\cite{AIKT99,AIKT02}, 
where the renormalization
factors for the four-quark operators relevant in the SM are evaluated.
 
This paper is organized as follows. In Sec.~\ref{sec:action} we introduce
the quark and gauge actions and the corresponding Feynman rules.
Section~\ref{sec:calculation} is devoted to explain 
our calculational procedure of
the renormalization factors for the complete set 
of $\Delta s=2$ four-quark operators. In Sec.~\ref{sec:susy}, 
with the use of our results,
we construct the renormalization factors for the SUSY operators 
in Eqs.(\ref{eq:op_susy1})$-$(\ref{eq:op_susy5}).
We briefly discuss the mean 
field improvement in Sec.~\ref{sec:mf}.
Our conclusions are summarized in Sec.~\ref{sec:conclusion}.

 The physical quantities are expressed in lattice units
and the lattice spacing $a$ is suppressed unless
necessary.
 We take SU($N$) gauge group with the gauge coupling $g$
and the second Casimir $C_F=(N^2-1)/(2N)$, while $N=3$
is specified in the numerical calculations. 

\section{Action and Feynman rules}
\label{sec:action}

We take Shamir's domain-wall fermion action\cite{Shamir93}
given by
\bea
S_{\rm DW} &=&
\sum_n\sum_{s=1}^{N_s}
 \Biggl[
\frac{1}{2}\sum_{\mu}\bigl[\bpsi(n)_s(-r+\gamma_\mu)U_\mu(n)\psi(n+\mu)_s
 +\bpsi(n)_s(-r-\gamma_\mu)U^\dagger_\mu(n-\mu)\psi(n-\mu)_s \bigl] 
 \nn\\&&
 +\frac{1}{2}\bigl[\bpsi(n)_s(1+\gamma_5)\psi(n)_{s+1}
 +\bpsi(n)_s(1-\gamma_5)\psi(n)_{s-1}\bigl]
 +(M-1+4r)\bpsi(n)_s\psi(n)_s
 \Biggl]
 \nn\\&&
 +m\sum_n\bigl[\bpsi(n)_{N_s}P_R\psi(n)_1+\bpsi(n)_1P_L\psi(n)_{N_s}\bigl],
\eea
where $n$ is a four-dimensional space-time coordinate and
 $s$ is a fifth-dimensional or ``flavor'' index bounded 
as $1\leqslant s\leqslant N_s$.
In this paper we conventinally take $N_s
 \rightarrow \infty$ limit to avoid complications
 arising from the finite $N_s$ such as mixing of
 operators with different chiralities.
The domain-wall height $M$ is a parameter of the theory
 which we set $0< M < 2$ to realize the massless
 fermion at tree level.
$P_{R/L}$ is a projection operator
 $P_{R/L}=(1\pm\gamma_5)/2$ and
 the Wilson parameter is set to $r=-1$.
The ``physical'' quark field are defined 
by the boundary fermions in the fifth dimensional space: 
\bea
 q(n) &=& P_R\psi(n)_1+P_L\psi(n)_{N_s},
\label{eq:q_dw} \\
 \bar{q}(n) &=& \bpsi(n)_{N_s}P_R+\bpsi(n)_1P_L,
\label{eq:qbar_dw}
\eea
whose mass is given by $m$. 
Our renormalization procedure is based on the
Green functions consisting of only the ``physical''
quark fields.

For the gauge part of the action we employ the
following form in four dimensions:
\bea
S_{\rm gluon}=\frac{1}{g^2}\Bigl[c_0\sum_{\rm plaquette}{\rm Tr} U_{\rm pl}
  +c_1\sum_{\rm rectangle}{\rm Tr}U_{\rm rtg}+c_2\sum_{\rm chair}{\rm
  Tr}U_{\rm chr} 
  +c_3\sum_{\rm parallelogram}{\rm Tr}U_{\rm plg}\Bigl],
\eea
where the first term represents the standard plaquette
action and the remaining terms are six-link loops formed
by a $1\times2$ rectangle, a bent $1\times2$ rectangle
(chair) and a 3-dimensional parallelogram.
The coefficients $c_0,\dots,c_3$ satisfy the
normalization condition
\bea
c_0+8c_1+16c_2+8c_3=1.
\eea
The RG improved gauge action is defined by setting the
parameters to the value suggested by an approximate
renormalization group analysis.
In the following we adopt $c_1=-0.331,c_2=c_3=0$ (Iwasaki)\cite{Iwasaki83}
for the RG improved gauge action
in addition to $c_1=c_2=c_3=0$ (plaquette).
With these choices of parameters the RG improved gauge
action is expected to realize smooth gauge field
fluctuations approximating those in the continuum limit
better than with the unimproved plaquette action.

Weak coupling perturbation theory is developed by expressing the link
variable in terms of the gauge potential
\bea
U_{x,\mu}=\exp(igA_\mu(x+\frac{1}{2}\hat{\mu}))
\label{eqn:link}
\eea
and expanding in terms of the gauge coupling.
We adopt a covariant gauge fixing with a gauge parameter $\alpha$ 
defined by  
\bea
S_{\rm GF} = \sum_x \frac{1}{2\alpha}
\left( \nabla_\mu A_\mu^a (x+\frac{1}{2}\hat{\mu}) \right)^2,
\eea
where $\nabla_\mu f_n\equiv (f_{n+\hat{\mu}}-f_n)$.

The free part of the gluon action takes the form in momentum space
\bea
S_0 = \frac{1}{2} \int_{-\pi}^{\pi}\frac{d^4k}{(2\pi)^4}
\sum_{\mu, \nu}
A_\mu^a(k) 
\left(G_{\mu\nu}(k)
-\left(1-\frac{1}{\alpha}\right)\hat{k}_\mu\hat{k}_\nu\right)
A_\nu^a(-k) , 
\eea
where
\bea
G_{\mu \nu}(k) = \hat{k}_\mu \hat{k}_\nu + \sum_\rho 
(\hat{k}_\rho \delta_{\mu \nu} - \hat{k}_\mu \delta_{\rho
\nu}) q_{\mu \rho} \hat{k}_\rho
\eea
with 
\bea
\hat{k}_\mu = 2 \sin\frac{k_\mu}{2}
\eea
and $q_{\mu \nu}$ is defined as
\bea
q_{\mu \nu} = (1-\delta_{\mu\nu})\left(1 - (c_1 - c_2 - c_3)
(\hat{k}_\mu^2 + \hat{k}_\nu^2) -(c_2 + c_3) \hat{k}^2\right) .
\eea
The gluon propagator can be written as 
\bea
D_{\mu \nu}(k) &=& (\hat{k}^2)^{-2} \left[
\hat{k}_\mu \hat{k}_\nu  + \sum_\sigma 
(\hat{k}_\sigma \delta_{\mu \nu} - \hat{k}_\nu \delta_{\mu
\sigma} ) \hat{k}_\sigma A_{\sigma \nu} \right]
-(1-\alpha) \frac{\hat{k}_\mu \hat{k}_\nu}{(\hat{k}^2)^2}
\\
& = & (\hat{k}^2)^{-2} \left[
(1 - A_{\mu \nu} )\hat{k}_\mu \hat{k}_\nu 
+ \delta_{\mu \nu} \sum_\sigma \hat{k}_\sigma^2
A_{\nu \sigma} \right]
-(1-\alpha) \frac{\hat{k}_\mu \hat{k}_\nu}{(\hat{k}^2)^2} , 
\eea
where $A_{\mu \nu}$ is a function of $q_{\mu \nu}$ and
$\hat{k}_\mu$, whose form we refer to the original 
literatures\cite{Iwasaki83,Weisz83}. 
In this paper we will adopt the Feynman gauge($\alpha=1$) without loss of
generality, since the renormalization factors for the composite
operators do not depend on the choice of the gauge fixing condition.

Quark-gluon vertices are also identical to those in the
$N_s$ flavor Wilson fermion.
We need only one gluon vertex for our present
calculation:
\bea
V_{1\mu}^A(k,p)_{st}=-igT^A\{\gamma_\mu\cos(-k_\mu/2+p_\mu/2)
    -ir\sin(-k_\mu/2+p_\mu/2)\}\delta_{st},
\eea
where $k$ and $p$ represent incoming momentum into the
vertex (see Fig.1 of Ref.\cite{AIKT98}) and
$T^A (A=1,\dots,N^2-1)$ is a generator of color SU($N$).

The fermion propagator originally takes $N_s\times N_s$
matrix form in $s$-flavor space.
In the present one-loop calculation,
however, we do not need the whole matrix elements because
we consider Green functions consisting of the physical
quark fields.
The relevant fermion propagators are restricted to the
following three types:
\bea
\langle q(-p)\bar{q}(p)\rangle&=& 
      \frac{-i\gamma_\mu\sin p_\mu+(1-We^{-\alpha})m}{-(1-e^\alpha W)
      +m^2(1-We^{-\alpha})}\equiv S_q(p),  \\
\langle q(-p)\bpsi(p,s)\rangle&=&
     \frac{1}{F}\bigl[i\gamma_\mu\sin p_\mu-m(1-We^{-\alpha})\bigl]
        (e^{-\alpha(N_s-s)}P_R+e^{-\alpha(s-1)}P_L) \nn \\&&
     +\frac{1}{F}\Bigl[m\bigl[i\gamma_\mu\sin
                     p_\mu-m(1-We^{-\alpha})\bigl]-F\Bigl]
        e^{-\alpha}(e^{-\alpha(s-1)}P_R+e^{-\alpha(N_s-s)}P_L),  \\
\langle \psi(-p,s)\bar{q}(p)\rangle
        &=&\frac{1}{F}(e^{-\alpha(N_s-s)}P_L+e^{-\alpha(s-1)}P_R)
               \bigl[i\gamma_\mu\sin  p_\mu-m(1-We^{-\alpha})\bigl]  \nn\\&&
         +\frac{1}{F} (e^{-\alpha(s-1)}P_L+e^{-\alpha(N_s-s)}P_R)e^{-\alpha}
            \Bigl[m\bigl[i\gamma_\mu\sin
                     p_\mu-m(1-We^{-\alpha})\bigl]-F\Bigl]
\eea
with
\bea
 W &=& 1-M-r\sum_\mu(1-\cos p_\mu), \\
\cosh (\alpha)&=&\frac{1+W^2+\sum_\mu\sin^2p_\mu}{2|W|} ,\\ 
 F &=& 1-e^\alpha W-m^2(1-We^{-\alpha}) ,
\eea
where the argument $p$ in the factors $\alpha$ and $W$ is suppressed.

In the perturbative calculation of Green functions
we assume that the
external quark momenta and masses are much
smaller than the lattice cutoff,
so that the external quark propagators can be expanded in
terms of them.
We have the following expressions as leading term of the
expansion:
\bea
\langle q\bar{q}\rangle_{\rm ext}(p) &=&\frac{1-w_0^2}{i\pslash+(1-w_0^2)m}, \\
\langle q\bpsi_s\rangle_{\rm ext}(p) &=&\langle q\bar{q}\rangle(p)
                    (w_0^{s-1}P_L+w_0^{N_s-s}P_R), \\
\langle \psi_s\bar{q}\rangle_{\rm ext}(p) &=&(w_0^{s-1}P_R+w_0^{N_s-s}P_L)\langle q\bar{q}\rangle(p),
\eea
where $w_0=1-M$.

\section{Renormalization factors for generic four-quark operators}
\label{sec:calculation} 
%
We consider the complete set of parity-conserving 
four-quark operators\cite{bk_w,DGMTV99}:
\bea
Q_1^{\pm} &\equiv&
   \mathcal{O}_{VV+AA}^\pm\equiv\mathcal{O}_{VV}^\pm+\mathcal{O}_{AA}^\pm,  
\label{eq:op_gen1}\\
Q_2^{\pm} &\equiv&
   \mathcal{O}_{VV-AA}^\pm\equiv\mathcal{O}_{VV}^\pm-\mathcal{O}_{AA}^\pm,  \label{eq:op_gen2}\\
Q_3^{\pm} &\equiv&
   \mathcal{O}_{SS-PP}^\pm\equiv\mathcal{O}_{SS}^\pm-\mathcal{O}_{PP}^\pm,  \label{eq:op_gen3} \\
Q_4^{\pm} &\equiv&
   \mathcal{O}_{SS+PP}^\pm\equiv\mathcal{O}_{SS}^\pm+\mathcal{O}_{PP}^\pm,  \label{eq:op_gen4}\\
Q_5^{\pm} &\equiv&
   \mathcal{O}_{TT}^\pm, \label{eq:op_gen5}
\eea
where
\bea
\mathcal{O}_{\Gamma\Gamma}^\pm=\frac{1}{2}
  \Bigl[(\bar{q}_1^a\Gamma q_2^a)(\bar{q}_3^b\Gamma q_4^b)\pm
     (\bar{q}_1^a\Gamma q_4^a)(\bar{q}_3^b\Gamma q_2^b)  \Bigl]
\label{eq:op_gen}
\eea
with
$\Gamma=\{\mathbf{1},\gamma_\mu,\sigma_{\mu\nu},\gamma_\mu\gamma_5,\gamma_5\}\equiv\{S,V,T,A,P\}$.
$a,b$ denote color indices and summation over them is assumed.
We should note that $q_i$ (i=1,2,3,4) are the physical quark fields 
defined by Eqs.(\ref{eq:q_dw}) and (\ref{eq:qbar_dw}). 
The parity-conserving parts of 
Eqs.(\ref{eq:op_susy1})$-$(\ref{eq:op_susy5}) are expressed as 
linear combinations of the above operators.
We find that Ref.\cite{bk_w} employs another choice of basis called 
Fierz eigenbasis, which is more convenient in considering 
the Fierz transformation of the four-quark operators.

We consider the following Green functions:
\bea
\langle Q^\pm_{i}\rangle_{\alpha\beta,\gamma\delta}^{ij,kl}&\equiv
\langle Q^\pm_{i}(q_1)_\alpha^i(\bar{q}_2)_\beta^j
(q_3)_\gamma^k(\bar{q}_4)_\delta^l\rangle, 
\eea
where $\alpha, \beta, \gamma, \delta$ 
are spinor indices and $i, j, k, l$ are color indices. 
It is convenient to decompose the above Green functions as
\bea
\langle\mathcal{O}^\pm_{\Gamma\Gamma}\rangle_{\alpha\beta,\gamma\delta}^{ij,kl}
&\equiv&
\langle\mathcal{O}^\pm_{\Gamma\Gamma}(q_1)_\alpha^i(q_2)_\beta^j
(q_3)_\gamma^k(q_4)_\delta^l\rangle \notag \\
&=&\frac{1}{2}\langle[(\bar{q}_1^a\Gamma q_2^a)(\bar{q}_3^b\Gamma q_4^b)\pm
(\bar{q}_1^a\Gamma q_4^a)(\bar{q}_3^b\Gamma q_2^b)]
(q_1)_\alpha^i(q_2)_\beta^j(q_3)_\gamma^k(q_4)_\delta^l\rangle \notag \\
&=&\frac{1}{2}\Bigl[\langle (q_1)_\alpha^i\bar{q}_1^a\Gamma q_2^a(\bar{q}_2)_\beta^j
(q_3)_\gamma^k\bar{q}_3^b\Gamma q_4^b(\bar{q}_4)_\delta^l\rangle
\mp
\langle (q_1)_\alpha^i\bar{q}_1^a\Gamma q_4^a(\bar{q}_4)_\delta^l
 (q_3)_\gamma^k\bar{q}_3^b\Gamma q_2^b(\bar{q}_2)_\beta^j
\rangle
\Bigl]. \label{matrixelements}
\eea
After trancating the external quark
propagators from $\langle Q^\pm_{i}\rangle$
with the multiplication of $i\pslash+(1-w_0^2)m$,
we obtain the vertex functions, which is written in the
following form up to the one-loop level:
\bea
(1-w^2_0)^4(\Lambda_i^\pm)_{\alpha\beta;\gamma\delta}^{ij;kl}
    =(1-w_0^2)^4(\Lambda^{(0)\pm}_i+\Lambda_i^{(1)\pm})^{ij;kl}_{\alpha\beta;\gamma\delta},
\eea
where the superscript $(n)$ refers to the $n$th loop
level and the subscript $i$ identifies the operators 
(\ref{eq:op_gen1})$-$(\ref{eq:op_gen5}).
The trivial factor $(1-w_0^2)^4$, which originates from the
external quark propagators, is factored out for convenience.
Since the renormalization factor does not depend on the
external momenta $p_i$, we suppress them.

The tree level vertex functions $\Lambda^{(0)\pm}_i$ are given by
\bea
i&=&1,\hspace{0.5cm} \frac{1}{2}[V\otimes V+A\otimes A 
    \mp(V\odot V+A\odot A)]_{\alpha\beta;\gamma\delta}[1\tilde{\otimes}1]^{ij;kl},  \\
i&=&2,\hspace{0.5cm} \frac{1}{2}[V\otimes V-A\otimes A 
    \mp(V\odot V-A\odot A)]_{\alpha\beta;\gamma\delta}[1\tilde{\otimes}1]^{ij;kl}, \\
i&=&3,\hspace{0.5cm} \frac{1}{2}[S\otimes S-P\otimes P 
     \mp(S\odot S-P\odot P)]_{\alpha\beta;\gamma\delta}[1\tilde{\otimes}1]^{ij;kl}, \\
i&=&4,\hspace{0.5cm} \frac{1}{2}[S\otimes S+P\otimes P 
     \mp(S\odot S+P\odot P)]_{\alpha\beta;\gamma\delta}[1\tilde{\otimes}1]^{ij;kl}, \\
i&=&5,\hspace{0.5cm} \frac{1}{2}[T\otimes T \mp T\odot T]_{\alpha\beta;\gamma\delta}[1\tilde{\otimes}1]^{ij;kl},
\eea
where $\otimes,\odot$ act on the Dirac spinor space
as
$[\Gamma\otimes\Gamma]_{\alpha\beta;\gamma\delta}
\equiv(\Gamma)_{\alpha\beta}(\Gamma)_{\gamma\delta}$,
$[\Gamma\odot\Gamma]_{\alpha\beta;\gamma\delta}
\equiv(\Gamma)_{\alpha\delta}(\Gamma)_{\gamma\beta}$,
and $\tilde{\otimes},\tilde{\odot}$ on the color space as
$[1\tilde{\otimes}1]^{ij:kl}\equiv\delta_{ij}\delta_{kl}$,
$[1\tilde{\odot}1]^{ij:kl}\equiv\delta_{il}\delta_{kj}$.

\begin{figure}
\begin{center}
\begin{picture}(420,450)(0,20)
\Text(60,60)[l]{${\rm a}^\prime$}
\Vertex(70,166){4.5}
\Vertex(70,154){4.5}
\ArrowLine(68,166)(8,226)
\ArrowLine(132,226)(72,166)
\ArrowLine(68,154)(8,94)
\ArrowLine(132,94)(72,154)
\Gluon(20,105)(120,105){5}{8}
%
\Text(60,260)[l]{a}
\Text(0,440)[l]{$\alpha,i$}
\Text(115,440)[l]{$\beta,j$}
\Text(0,285)[l]{$\gamma,k$}
\Text(115,285)[l]{$\delta,l$}
\Vertex(70,366){4.5}
\Vertex(70,354){4.5}
\ArrowLine(68,366)(8,426)
\ArrowLine(132,426)(72,366)
\ArrowLine(68,354)(8,294)
\ArrowLine(132,294)(72,354)
\Gluon(20,415)(120,415){-5}{8}
%
\Text(200,60)[l]{${\rm b}^\prime$}
\Vertex(210,166){4.5}
\Vertex(210,154){4.5}
\ArrowLine(208,166)(148,226)
\ArrowLine(272,226)(212,166)
\ArrowLine(208,154)(148,94)
\ArrowLine(272,94)(212,154)
\Gluon(260,213)(260,107){-5}{8}
%
\Text(200,260)[l]{b}
\Vertex(210,366){4.5}
\Vertex(210,354){4.5}
\ArrowLine(208,366)(148,426)
\ArrowLine(272,426)(212,366)
\ArrowLine(208,354)(148,294)
\ArrowLine(272,294)(212,354)
\Gluon(160,413)(160,307){5}{8}
%
\Text(340,60)[l]{${\rm c}^\prime$}
\Vertex(350,166){4.5}
\Vertex(350,154){4.5}
\ArrowLine(348,166)(288,226)
\ArrowLine(412,226)(352,166)
\ArrowLine(348,154)(288,94)
\ArrowLine(412,94)(352,154)
\GlueArc(350,160)(30,51,230){-5}{7}
%
\Text(340,260)[l]{c}
\Vertex(350,366){4.5}
\Vertex(350,354){4.5}
\ArrowLine(348,366)(288,426)
\ArrowLine(412,426)(352,366)
\ArrowLine(348,354)(288,294)
\ArrowLine(412,294)(352,354)
\GlueArc(350,360)(30,-50,130){-5}{7}
\end{picture}
\end{center}
\caption{One-loop vertex corrections for the four-quark operator.
$\alpha,\beta,\gamma,\delta$ and $i,j,k,l$ label Dirac and color
indices respectively.}
\label{fig:vc_4}
\end{figure}
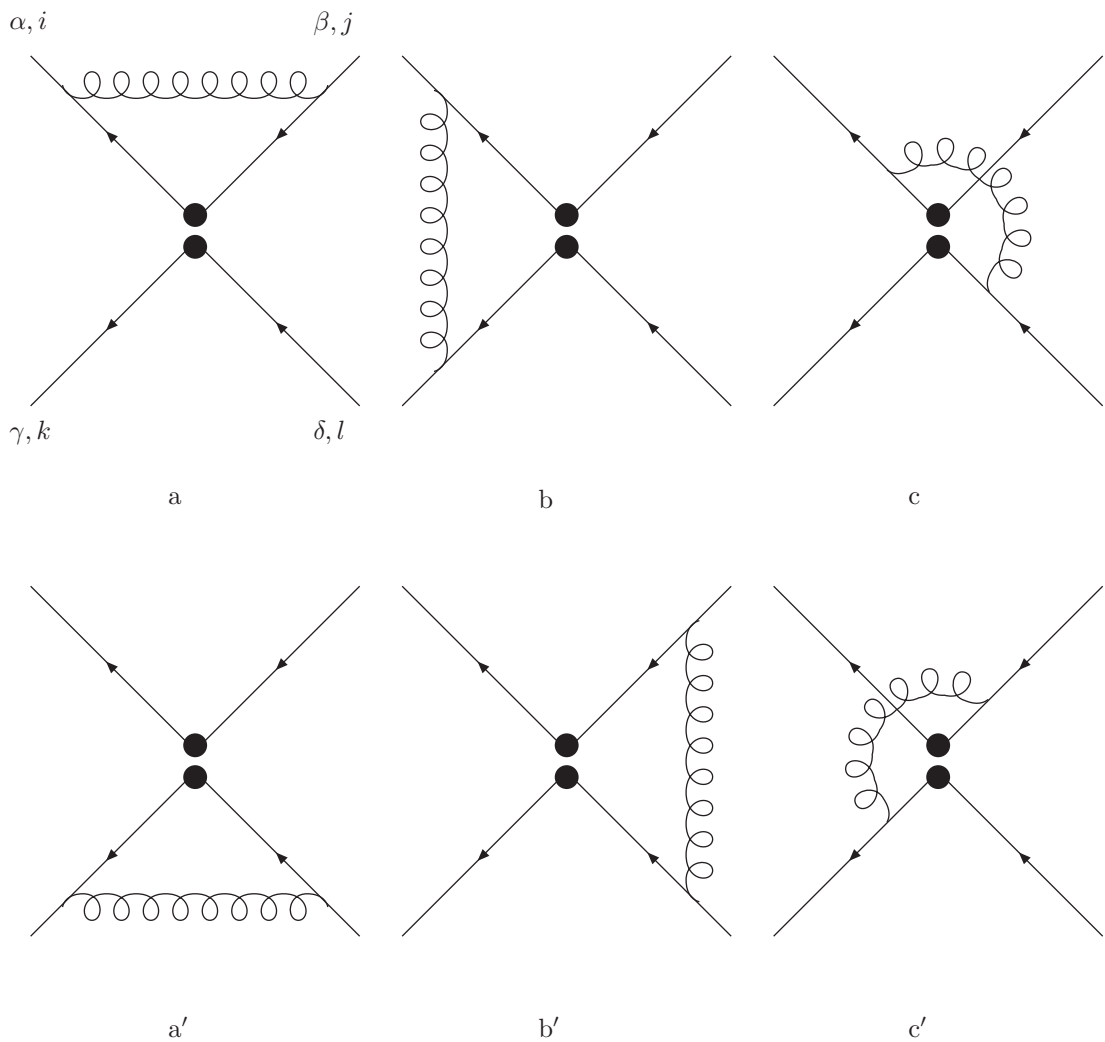

Now let us consider the one-loop vertex corrections depicted
by six diagrams in Fig.\ref{fig:vc_4}.
Their total contribution yields the vertex function at one-loop level:
\bea
i&=&1,\hspace{0.5cm} \Lambda^{(1)\pm}_1
     =\int_{-\pi}^{\pi}\frac{d^4k}{(2\pi)^4}\Bigl[\bigl(I_{VV}^a+I_{AA}^a\bigl)+,\dots,+\bigl(I_{VV}^{c\prime}+I_{AA}^{c\prime}\bigl)\Bigl], \\
i&=&2,\hspace{0.5cm} \Lambda^{(1)\pm}_2
     =\int_{-\pi}^{\pi}\frac{d^4k}{(2\pi)^4}\Bigl[\bigl(I_{VV}^a-I_{AA}^a\bigl)+,\dots,+\bigl(I_{VV}^{c\prime}-I_{AA}^{c\prime}\bigl)\Bigl], \\
i&=&3,\hspace{0.5cm} \Lambda^{(1)\pm}_3
     =\int_{-\pi}^{\pi}\frac{d^4k}{(2\pi)^4}\Bigl[\bigl(I_{SS}^a-I_{PP}^a\bigl)+,\dots,+\bigl(I_{SS}^{c\prime}-I_{PP}^{c\prime}\bigl)\Bigl], \\
i&=&4,\hspace{0.5cm} \Lambda^{(1)\pm}_4
     =\int_{-\pi}^{\pi}\frac{d^4k}{(2\pi)^4}\Bigl[\bigl(I_{SS}^a+I_{PP}^a\bigl)+,\dots,+\bigl(I_{SS}^{c\prime}+I_{PP}^{c\prime}\bigl)\Bigl], \\
i&=&5,\hspace{0.5cm} \Lambda^{(1)\pm}_5
     =\int_{-\pi}^{\pi}\frac{d^4k}{(2\pi)^4}\Bigl[I_{TT}^{a}+,\dots,+I_{TT}^{c\prime}\Bigl],
\eea
where
\bea
I_{\Gamma\Gamma}^a&=& \frac{1}{2}g^2  (T^AT^A\tilde{\otimes}1)
   K\Bigl[\{\gamma_\alpha\gamma_\beta\Gamma\gamma_\beta\gamma_\alpha\otimes\Gamma
     \bigl(\frac{1}{4}A_{\alpha\beta}\sin^2k_\alpha\sin^2k_\beta
    +\sum_{\sigma}A_{\alpha\sigma}\sin^2\frac{k_\sigma}{2}\cos^2\frac{k_\alpha}{2}\sin^2k_\beta\bigl)\nn\\
    &+&\Gamma\otimes\Gamma\bigl(4\Delta_3^2-\frac{1}{4}\sum_{\sigma}\Delta_{1,1}^\alpha
              +\frac{1}{4}\sum_\alpha\sin^4k_\alpha+T\bigl)\Bigl\}
             \mp\{\otimes\leftrightarrow\odot\}],  \label{iga}\\
I_{\Gamma\Gamma}^b&=& \frac{1}{2}g^2  (T^A\tilde{\otimes}T^A)
    K \Bigl[\{\gamma_\alpha\gamma_\beta\Gamma\otimes\Gamma\gamma_\beta\gamma_\alpha
      \bigl(\frac{1}{4}A_{\alpha\beta}\sin^2k_\alpha\sin^2k_\beta
    +\sum_{\sigma}A_{\alpha\beta}\sin^2\frac{k_\sigma}{2}\cos^2\frac{k_\alpha}{2}\sin^2k_\beta\bigl)\nn\\
   &+&\Gamma\otimes\Gamma\bigl(4\Delta_3^2-\frac{1}{4}\sum_\alpha\Delta_{1,1}^\alpha
             +\frac{1}{4}\sum_\sigma\sin^4k_\sigma+T\bigl) \Bigl\}
            \mp\{\otimes\leftrightarrow\odot\}], \label{igb} \\
I_{\Gamma\Gamma}^c&=& -\frac{1}{2}g^2 (T^A\tilde{\otimes}T^A)
    K \Bigl[\{\gamma_\alpha\gamma_\beta\Gamma\otimes\gamma_\alpha\gamma_\beta\Gamma 
      \bigl(\frac{1}{4}A_{\alpha\beta}\sin^2k_\alpha\sin^2k_\beta
     +\sum_{\sigma}A_{\alpha\sigma}\sin^2\frac{k_\sigma}{2}\cos^2\frac{k_\alpha}{2}\sin^2k_\beta\bigl)\nn\\
   &+&\Gamma\otimes\Gamma\bigl(4\Delta_3^2-\frac{1}{4}\sum_\sigma\Delta_{1,1}^\sigma
            +\frac{1}{4}\sum_\sigma\sin^2k_\sigma+T\bigl) \Bigl\}
            \mp\{\otimes\leftrightarrow\odot\}] \label{igc}
\eea
with
\bea
   K&\equiv&\frac{4}{\left(4\sum_{\mu}\sin^2\frac{k_{\mu}}{2}\right)^2}
              \frac{1}{\tilde{F}_0^2\tilde{F}^2}, \\
   \tilde{F}_0&\equiv& e^\alpha-w_0, \\
   \tilde{F} &\equiv& e^{-\alpha}-W,\\
   T &\equiv& r^2\Delta_1^2\tilde{F}^2+4r\Delta_3\Delta_1\tilde{F},\\
\Delta_3 &=& \frac{1}{4}\sum_\mu\sin^2k_\mu, \\ 
\Delta_1 &=& \sum_\mu\sin^2\frac{k_\mu}{2},  \\
\Delta_{1,1}^\mu &=&
\sum_\nu(\delta_{\mu\nu}+A_{\mu\nu})\sin^2k_\mu\sin^2k_\nu.
\eea
Summation over repeated indices is assumed.
The above expressions are obtained with the aid of
useful formula presented in Refs.\cite{AIKT98,AIKT99,AIKT02}.  
We should note that the other three contributions
$I_{\Gamma\Gamma}^{a^\prime,b^\prime,c^\prime}$ 
from Figs.~1$a^\prime$, 1$b^\prime$, 1$c^\prime$ are
equal to $I_{\Gamma\Gamma}^{a,b,c}$ respectively. 
After all, the total contribution becomes
\bea
\bigl(I_{VV}^a+I_{AA}^a\bigl)+,\dots,+\bigl(I_{VV}^{c\prime}+I_{AA}^{c\prime}\bigl) &=&
      g^2[V\otimes V+A\otimes A\mp(V\odot V+A\odot A)]_{\alpha\beta:\gamma\delta}
      [1\tilde{\otimes}1]^{ij:kl} \notag \\ &&
      \times K[(\frac{1}{N}\mp1)A_{SP}+(2C_F-\frac{1}{N}\pm1)A_{VA}+2C_FT], \\
\bigl(I_{VV}^a-I_{AA}^a\bigl)+,\dots,+\bigl(I_{VV}^{c\prime}-I_{AA}^{c\prime}\bigl) &=&
       g^2[V\otimes V-A\otimes A\mp(V\odot V-A\odot A)]_{\alpha\beta:\gamma\delta}
          [1\tilde{\otimes}1]^{ij:kl}  \notag \\&& 
          \times K[-\frac{1}{N}A_{SP}+(2C_F+\frac{1}{N})A_{VA}+2C_FT] \notag \\
    &&+g^2[S\otimes S-P\otimes P\mp(S\odot S-P\odot P)]_{\alpha\beta:\gamma\delta}
          [1\tilde{\otimes}1]^{ij:kl}  \notag \\&&
          \times K[\mp2 A_{SP}\pm 2A_{VA}], \\
\bigl(I_{SS}^a-I_{PP}^a\bigl)+,\dots,+\bigl(I_{SS}^{c\prime}-I_{PP}^{c\prime}\bigl)&=& 
       g^2[S\otimes S-P\otimes P\mp(S\odot S-P\odot P)]_{\alpha\beta:\gamma\delta} 
          [1\tilde{\otimes}1]^{ij:kl}\notag \\&&
          \times  K[2C_FA_{SP}+2C_FT],\\
\bigl(I_{SS}^a+I_{PP}^a\bigl)+,\dots,+\bigl(I_{SS}^{c\prime}+I_{PP}^{c\prime}\bigl)&=& 
       g^2[S\otimes S+P\otimes P\mp(S\odot S+P\odot P)]_{\alpha\beta:\gamma\delta}
          [1\tilde{\otimes}1]^{ij:kl} \notag \\ &&
          \times K[(2C_F\mp1)A_{SP}\pm A_{VA}+2C_FT] \notag \\
    &&+g^2(1-w_0^2)^2([T\otimes T\mp T\otimes T]_{\alpha\beta:\gamma\delta}
          [1\tilde{\otimes}1]^{ij:kl} \notag \\&&
      \times K[(\frac{2}{3N}\mp\frac{1}{3})A_{SP}+(-\frac{2}{3N}\pm\frac{1}{3})A_{VA})],  \\
I_{TT}^{a}+,\dots,+I_{TT}^{c\prime} &=& 
       g^2[T\otimes T\mp T\odot T]_{\alpha\beta:\gamma\delta}
       [1\tilde{\otimes}1]^{ij:kl} \notag \\&&
        \times K[(-\frac{2C_F}{3}\mp1)A_{SP}
            +(\frac{8C_F}{3}\pm1)A_{VA}+2C_FT]\notag  \\
    &&+g^2[S\otimes S+P\otimes P\mp(S\odot S+ P\odot P)]_{\alpha\beta:\gamma\delta}  
       [1\tilde{\otimes}1]^{ij:kl} \notag \\&&
      \times K(\frac{2}{N}\pm1)[A_{SP}-A_{VA}], 
\eea
where we use the Fierz rearrangements:
\bea
\bigl[S\otimes S\bigl]\bigl[1\tilde{\odot}1\bigl] &=& 
        \frac{1}{4}\bigl[S\odot S+V\odot V-T\odot T-A\odot A+P\odot P\bigl]
                                      \bigl[1\tilde{\otimes}1\bigl], \\
\bigl[V\otimes V\bigl]\bigl[1\tilde{\odot}1\bigl] &=& 
        \frac{1}{4}\bigl[4S\odot S-2V\odot V-2A\odot A-4P\odot P\bigl]
                                      \bigl[1\tilde{\otimes}1\bigl], \\
\bigl[T\otimes T\bigl]\bigl[1\tilde{\odot}1\bigl] &=& 
        \frac{1}{4}\bigl[-6S\odot S-2T\odot T-6P\odot P\bigl]
                                      \bigl[1\tilde{\otimes}1\bigl], \\
\bigl[A\otimes A\bigl]\bigl[1\tilde{\odot}1\bigl] &=& 
        \frac{1}{4}\bigl[-4S\odot S-2V\odot V-2A\odot A+4P\odot P\bigl]
                                      \bigl[1\tilde{\otimes}1\bigl], \\
\bigl[P\otimes P\bigl]\bigl[1\tilde{\odot}1\bigl] &=& 
        \frac{1}{4}\bigl[S\odot S-V\odot V-T\odot T+A\odot A+P\odot P\bigl]
                                      \bigl[1\tilde{\otimes}1\bigl]. 
\eea
Note that these formula do not include
Fermi statistics.

Comparing the one-loop results to the tree level ones we obtain
\bea
\Lambda_1^{\pm} &=& \Bigl[1+g^2\frac{N\mp1}{N}\{\mp\langle A_{SP}\rangle
      +(N\pm2)\langle A_{VA}\rangle+(N\pm1)\langle
 T\rangle\}\Bigl]\Lambda_1^{(0)\pm}, \label{lam1}\\
\Lambda_2^{\pm} &=& \Bigl[1+g^2\{-\frac{1}{N}\langle A_{SP}\rangle
      +(2C_F+\frac{1}{N})\langle A_{VA}\rangle
      +2C_F\langle T\rangle\}\Bigl]\Lambda_2^{(0)\pm} \notag \\&&
      +g^2\{\mp2\langle A_{SP}\rangle\pm2\langle  A_{VA}\rangle\}\Lambda_3^{(0)\pm}, \label{lam2}\\ 
\Lambda_3^{\pm} &=& \Bigl[1+g^22C_F\{\langle A_{SP}\rangle
      +\langle T\rangle \}\Bigl]\Lambda_3^{(0)\pm}, \label{lam3}\\
\Lambda_4^{\pm} &=& \Bigl[1+g^2\frac{1}{N}\{(N^2\mp N-1)\langle A_{SP}\rangle
      \pm N\langle A_{VA}\rangle+(N^2-1)\langle T\rangle \}\Bigl]\Lambda_4^{(0)\pm}\notag \\&&
       +g^2\frac{2\mp N}{3N}\{\langle A_{SP}\rangle
       -\langle A_{VA}\rangle\}\Lambda_5^{(0)\pm}, \label{lam4} \\
\Lambda_5^{\pm} &=&
\Bigl[1+g^2\{(-\frac{2C_F}{3}\mp1)\langle A_{SP}\rangle
      +(\frac{8C_F}{3}\pm1)\langle A_{VA}\rangle
      +2C_F\langle T\rangle \}\Bigl]\Lambda_5^{(0)\pm} \notag \\&&
      +g^2(\frac{2}{N}\pm1)\{\langle A_{SP}\rangle-\langle A_{VA}\rangle\}
                                                        \Lambda_4^{(0)\pm} \label{lam5}
\eea
with
\bea
\langle X\rangle=\int_{-\pi}^\pi \frac{d^4k}{(2\pi)^4}K(k)X(k)
\eea
for $X=T,A_{VA},A_{SP}$.
We remark that $C_F\langle T+A_{VA}\rangle$ and
$C_F\langle T+A_{SP}\rangle$ correspond to the one-loop
vertex corrections to the (axial) vector and the
(pseudo) scalar density.
$A_{VA}$ and $A_{SP}$ are expressed as follows:
\bea
A_{VA} &=& 4\Delta_3^2-\Delta_{1,1}^\mu+4s_\mu^2\Delta_{1,0}^\mu, \\ 
A_{SP} &=& 16\Delta_3\Delta_{1,0}^\mu
\eea
with
\bea
\Delta_{1,0}^\mu &=&
\sum_\nu(\delta_{\mu\nu}+A_{\mu\nu})\cos^2\frac{k_\mu}{2}\sin^2\frac{k_\nu}{2}. 
\eea
The expressions of Eqs.\eqref{lam1} and \eqref{lam3}
show  important properties of $Q_1$ and $Q_3$ in DWQCD
formalism:
the one-loop vertex corrections are multiplicative.
This is contrary to the Wilson case, in which the mixing
operators with different chiralities appears at the
one-loop level.
The results for $Q_1^{\pm}$ in Eq.\eqref{lam1} 
are already obtained in Ref.~\cite{AIKT99}.

Combining the contribution from the quark self-energy
evaluated in Ref.~\cite{AIKT98} and the vertex corrections, 
we obtain the lattice renormalization factors:
\bea
Z_{ij}^{\pm{\rm lat}} =(1-w_0^2)^2Z_w^2Z_2^2V_{ij}^\pm, 
\eea
where 
\bea
Z_2 &=& 1+\frac{g^2}{16\pi^2}C_F[\log(\lambda a)^2+\Sigma_1], \\
Z_w &=& 1-\frac{2w_0}{1-w_0^2}\frac{g^2C_F}{16\pi^2}\Sigma_3, \\
V_{ij}^\pm &=& \delta_{ij}+\frac{g^2}{16\pi^2}\bigl[\gamma_{ij}^\pm\log(\lambda a)^2+v_{ij}^\pm\bigl].
\eea\
$\lambda$ is the fictitious gluon mass introduced 
to regularize the infrared divergences.
$\Sigma_1$ and $\Sigma_3$ are finite parts of the
renormalization factors of quark wave function and
overall factor $(1-w_0^2)$, whose numerical value is
given in Ref.\cite{AIKT02}.
The matrix $v_{ij}^\pm$ is expressed as
\bea
v_{ij}^\pm=\left(
         \begin{array}{@{\,}ccccc@{\,}}
           v_{11}^\pm&0&0&0&0 \\
           0&v_{22}^\pm& v_{23}^\pm&0&0 \\
           0&0     & v_{33}^\pm&0&0 \\
           0&0&0&v_{44}^\pm&v_{45}^\pm \\
           0&0&0&v_{54}^\pm&v_{55}^\pm \\
         \end{array}
         \right),
\eea
whose components are
\bea
v_{11}^+ &=& \frac{16\pi^2(N-1)}{N}\bigl[-\langle\langle A_{SP}\rangle\rangle
          +(N+2)\langle\langle A_{VA}\rangle\rangle+(N+1)\langle T\rangle\bigl]
          +\gamma_{11}^+\log\pi^2, \label{v1p}\\
v_{11}^- &=& \frac{16\pi^2(N+1)}{N}\bigl[\langle\langle A_{SP}\rangle\rangle
          +(N-2)\langle\langle A_{VA}\rangle\rangle+(N-1)\langle T\rangle\bigl]
          +\gamma_{11}^-\log\pi^2,  \label{v1m}\\
v_{22}^\pm &=& \frac{16\pi^2}{N}\bigl[-\langle\langle A_{SP}\rangle\rangle
          +N^2\langle\langle A_{VA}\rangle\rangle+(N^2-1)\langle T\rangle\bigl]
          +\gamma_{22}^\pm\log\pi^2, \\
v_{23}^\pm &=& \mp16\pi^2\bigl[2\langle\langle A_{SP}\rangle\rangle
          -2\langle\langle A_{VA}\rangle\rangle\bigl]
          +\gamma_{23}^\pm\log\pi^2, \\
v_{33}^\pm &=& \frac{16\pi^2(N+1)(N-1)}{N}\bigl[\langle\langle A_{SP}\rangle\rangle
          +\langle T\rangle\bigl]  
          +\gamma_{33}^\pm\log\pi^2, \\
v_{44}^\pm &=& \frac{16\pi^2}{N}\bigl[(N^2\mp N-1)\langle\langle A_{SP}\rangle\rangle
          \pm N\langle\langle A_{VA}\rangle\rangle+(N^2-1)\langle T\rangle\bigl]
          +\gamma_{44}^\pm\log\pi^2, \\
v_{45}^\pm &=& \frac{16\pi^2(2\mp N)}{3N}\bigl[\langle\langle A_{SP}\rangle\rangle
          -\langle\langle A_{VA}\rangle\rangle\bigl]
          +\gamma_{45}^\pm\log\pi^2, \\
v_{55}^\pm &=& \frac{16\pi^2}{N}\bigl[\frac{-N^2\mp3N+1}{3}\langle\langle A_{SP}\rangle\rangle
          +\frac{4N^2\pm3N-4}{3}\langle\langle A_{VA}\rangle\rangle+(N^2-1)\langle T\rangle\bigl]
          +\gamma_{55}^\pm\log\pi^2, \\
v_{54}^\pm &=& \frac{16\pi^2(2\pm N)}{N}\bigl[\langle\langle A_{SP}\rangle\rangle
          -\langle\langle A_{VA}\rangle\rangle\bigl]
          +\gamma_{54}^\pm\log\pi^2
\eea
with
\bea
\gamma_{ij}^{\pm} &=& \left(
         \begin{array}{@{\,}ccccc@{\,}}
           \gamma_{11}^\pm &0&0&0&0 \\
           0&\gamma_{22}^\pm& \gamma_{23}^\pm&0&0 \\
           0&0& \gamma_{33}^\pm&0&0 \\
           0&0&0&\gamma_{44}^\pm&\gamma_{45}^\pm \\
           0&0&0&\gamma_{54}^\pm&\gamma_{55}^\pm \\
         \end{array}
         \right) \notag \\
       &=&
        \left(
         \begin{array}{@{\,}ccccc@{\,}}
           \frac{(N\mp1)(N\mp2)}{N} &0&0&0&0 \\
           0&\frac{(N+2)(N-2)}{N}&\mp6&0&0 \\
           0&0& 4\frac{(N+1)(N-1)}{N}&0&0 \\
           0&0&0&\frac{4N^2\mp3N-4}{N}&\frac{2\mp N}{N} \\
           0&0&0&\frac{6\pm3N}{N}&\mp3 \\
         \end{array}
         \right). \label{anomaldim}
\eea
The infrared singularity in $\langle A_X\rangle$ is
subtracted as 
\bea
\langle\langle A_X\rangle\rangle =\int_{-\pi}^\pi\frac{d^4k}{(2\pi)^4}
        \Biggl[K(k)A_X(k)-c_X\frac{\theta(\pi^2-k^2)}{(k^2)^2}\Biggl],
\eea
where $c_{SP}=4$ and $c_{VA}=1$.

The lattice operators and the continuum ones
defined in the $\msbar$ scheme with the naive dimensional 
reguralization (NDR) are related as
\bea
Q_i^{\pm\msbar}(\mu) &=& \frac{1}{(1-\omega^2_0)^2Z_\omega^2}Z_{ij}^\pm(\mu
 a)Q_j^{\pm{\rm lat}}(1/a) 
\eea
with
\bea
Z_{ij}^\pm(\mu a) &\equiv& \frac{(Z_2^{\msbar})^2}{Z_2^2}\frac{V_{ij}^{\pm\msbar}}{V_{ij}^{\pm}}
\notag\\
  &=&
  \delta_{ij}+\frac{g^2}{16\pi^2}\left[(\gamma_{ij}-2C_F\delta_{ij})\log(\mu a)^2+z_{ij}^\pm\right]
\label{eq:zij_def}\\
  z_{ij}^\pm &=&
v_{ij}^{\pm\msbar}-v_{ij}^\pm+\delta_{ij}2C_F(\Sigma_1^{\msbar}-\Sigma_1) \\
\Sigma_1^{\msbar}({\rm NDR}) &=& \frac{1}{2},
\hspace{1cm}
v_{11}^{+\msbar}({\rm NDR})=-5,
\hspace{1cm}
v_{11}^{-\msbar}({\rm NDR})=6 \\
v_{ij}^{\pm\msbar}({\rm NDR}) &=& \left(
         \begin{array}{@{\,}cccc@{\,}}
           v_{22}^{\pm\msbar}& v_{23}^{\pm\msbar}&0&0 \\
           v_{32}^{\pm\msbar}& v_{33}^{\pm\msbar}&0&0 \\
           0&0&v_{44}^{\pm\msbar}&v_{45}^{\pm\msbar} \\
           0&0&v_{54}^{\pm\msbar}&v_{55}^{\pm\msbar} \\
         \end{array}
         \right) 
       = \left(
         \begin{array}{@{\,}cccc@{\,}}
           -\frac{7}{6}& -1&0&0 \\
           \pm\frac{3}{2}& \frac{19}{3}&0&0 \\
           0&0& \frac{17}{3}\mp 1 & \frac{1}{3}\mp1 \\
           0&0& 1  & \frac{1}{3}\mp{2} \\
         \end{array}
         \right)
\eea
where 
$v_{ij}^{\pm\msbar}$ are for the case of $N=3$.
The continuum counterparts of the wave-function renormalization factor
and the vertex corrections 
are evaluated by employing the same gauge fixing condition
and the same infrared regulator as the lattice case.
Here it should be noted that the one-loop vertex corrections require to
specify the evanescent operators, which originates from the property
that the Fierz transformation cannot be defined in the NDR scheme\cite{BW90}.
We employ the following evanescent operators:
\bea
E_{VV}^{\rm NDR}&=&\gamma_\alpha\gamma_\beta V\otimes V\gamma_\alpha\gamma_\beta
-\{(D^2-2D+2)V\otimes V+2(1-D)A\otimes A\},\\
E_{AA}^{\rm NDR}&=&\gamma_\alpha\gamma_\beta A\otimes A\gamma_\alpha\gamma_\beta
-\{2(1-D)V\otimes V+(D^2-2D+2)A\otimes A\},\\
E_{SS}^{\rm NDR}&=&\gamma_\alpha\gamma_\beta S\otimes S\gamma_\alpha\gamma_\beta
-\{(D^2-5D+8)S\otimes S+(4-D)P\otimes P+(2-D)T\otimes T\},\\
E_{PP}^{\rm NDR}&=&\gamma_\alpha\gamma_\beta P\otimes P\gamma_\alpha\gamma_\beta
-\{(4-D)S\otimes S+(D^2-5D+8)P\otimes P+(2-D)T\otimes T\},\\
E_{TT}^{\rm NDR}&=&\gamma_\alpha\gamma_\beta T\otimes T\gamma_\alpha\gamma_\beta
-\{6(2-D)S\otimes S+6(2-D)P\otimes P+(D^2-2D+4)T\otimes T\},
\eea
where $D$ is the reduced space-time dimension.

Numerical values for $v_{ij}$ and $z_{ij}$ are evaluated 
by momentum integration, which is performed by a mode sum 
for a periodic box of a
size $L^4$ after transforming the momentum variable
through $k_\mu=q_\mu-\sin q_\mu$.
We employ the size $L=64$ for integrals,
and numerical error is estimated by varying $L$ from
64 to 60. The results are presented in Table
I,$\dots$,VI for the plaquette gauge action and in Table
VII,$\dots$,XII for the Iwasaki gauge action as a
function of $M$.
These values are evaluated with $N=3$.

\section{Renormalization factors for SUSY operators} 
\label{sec:susy}
%
Let us consider the parity-conserving parts of the SUSY operators
in Eqs.(\ref{eq:op_susy1})$-$(\ref{eq:op_susy5}), 
which are relevant for $K^0-\kzbar$ mixing.
They are related to the operators 
of Eqs.(\ref{eq:op_gen1})$-$(\ref{eq:op_gen5}) as
\bea
{\cal O}_1 &=& Q_1^+, \\
{\cal O}_2 &=& Q_4^+, \\
{\cal O}_3 &=& -\frac{1}{2}(Q_4^+-Q_5^+), \\
{\cal O}_4 &=& Q_3^+, \\
{\cal O}_5 &=& -\frac{1}{2}Q_2^+,
\eea
where we take $q_1=q_3=s$ and $q_2=q_4=d$ in Eq.(\ref{eq:op_gen}).
With the use of these relations, 
the renormalization factors for ${\cal O}_i$ are expressed as
\bea
{\cal O}_1^{\msbar}(\mu) &=& \frac{1}{(1-\omega^2_0)^2Z_\omega^2}
[Z_{11}^+(\mu a){\cal O}_1^{{\rm lat}}(1/a)], \\
{\cal O}_2^{\msbar}(\mu) &=& \frac{1}{(1-\omega^2_0)^2Z_\omega^2}
[(Z_{44}^+(\mu a)+Z_{45}^+(\mu a)){\cal O}_2^{{\rm lat}}(1/a)
+2Z_{45}^+(\mu a){\cal O}_3^{{\rm lat}}(1/a)], \\
{\cal O}_3^{\msbar}(\mu) &=& \frac{1}{(1-\omega^2_0)^2Z_\omega^2}
[(Z_{55}^+(\mu a)-Z_{45}^+(\mu a)){\cal O}_3^{{\rm lat}}(1/a)\nn\\
&&-\frac{1}{2}(Z_{44}^+(\mu a)+Z_{45}^+(\mu a)
-Z_{54}^+(\mu a)-Z_{55}^+(\mu a)){\cal O}_2^{{\rm lat}}(1/a)], \\
{\cal O}_4^{\msbar}(\mu) &=& \frac{1}{(1-\omega^2_0)^2Z_\omega^2}
[Z_{33}^+(\mu a){\cal O}_4^{{\rm lat}}(1/a)
-2Z_{32}^+(\mu a){\cal O}_5^{{\rm lat}}(1/a)], \\
{\cal O}_5^{\msbar}(\mu) &=& \frac{1}{(1-\omega^2_0)^2Z_\omega^2}
[Z_{22}^+(\mu a){\cal O}_5^{{\rm lat}}(1/a)
-\frac{1}{2}Z_{23}^+(\mu a){\cal O}_4^{{\rm lat}}(1/a)], 
\eea
where $Z_{ij}^{+}$ are given in Eq.(\ref{eq:zij_def}). 

%
\section{Mean field improvement}
\label{sec:mf}
%
Let us briefly explain the mean-field improvement on the
renormalization factors of the four-quark operators in
Eqs.(\ref{eq:op_gen1})$-$(\ref{eq:op_gen5}).
We follow the discussion in Sec.VI of Ref.\cite{AIKT02}.
The renormalization factors with the mean field improvement 
for the four-quark operators are given by
\bea
Q_i^{\pm\msbar}(\mu) &=& \frac{u^2}{(1-\omega^2_0)^2Z_\omega^2}Z_{ij}^\pm(\mu
 a)Q_j^{\pm{\rm lat}}(1/a), \\ 
Z_{ij}^{\pm}&=&
  \delta_{ij}+\frac{g^2}{16\pi^2}\left[(\gamma_{ij}-2C_F\delta_{ij})\log(\mu a)^2+z_{ij}^{\pm{\rm MF}}\right], \\
z_{ij}^{\pm \rm MF} &=& 
v_{ij}^{\pm\msbar}-v_{ij}^\pm+\delta_{ij}2C_F(\Sigma_1^{\msbar}-\Sigma_1)
+\delta_{ij}16\pi^2C_FT_{MF},
\label{susybpmf}
\eea
where $T_{\rm MF}$ is the one-loop correction to the mean field factor 
defined by
\be
u=1-g^2 C_F\frac{T_{\rm MF}}{2}+\cdots 
\ee
and $w_0=1-{\tilde M}$ with ${\tilde M}=M-4(1-u)$.
Note that the overall factor $(1-\omega^2_0)^2Z_\omega^2$ is 
mean-field improved following the description in Ref.\cite{AIKT02}.

Now it is instructive to evaluate the magnitude of 
the renormalization factors in current representative simulations.
We take Ref.\cite{Noaki03} as an example, where 
a quenched simulation of domain-wall QCD was made at
$\beta=6.0$ for the plaquette gauge action with $M=1.8$ and
$\beta=2.6$ for the Iwasaki gauge action with $M=1.8$.
The mean field improved $\msbar$ coupling $g^2_{\msbar}(1/a)$
at the scale $\mu=1/a$ is obtained from
\bea
\frac{1}{g^2_{\msbar}}(1/a)=P\frac{\beta}{6}+d_g+c_p
\eea
for the plaquette gauge action, while 
\bea
\frac{1}{g^2_{\msbar}}(1/a)=(c_0 P+8c_1R )\frac{\beta}{6}+d_g
+c_0\cdot c_p+8c_1\cdot c_{R1}
\eea  
for the Iwasaki gauge action,
where the values of $d_g$ and $c_p$ are listed in Table XVI of
Ref.\cite{AIKT02}. $P$ denotes the expectation value of the plaquette
and $R$ for the $1\times 2$ rectangular.
Their values are taken from Ref.\cite{bk_cppacs}.
The domain-wall height $M=1.8$ 
is replaced with  
${\tilde M}=1.3112$ (plaquette) and ${\tilde M}=1.4198$ (Iwasaki), 
respectively, according to ${\tilde M}=M-4(1-u)$ with $u=P^{1/4}$.
The mean field improved values for the renormalization factors $Z_{ij}^{\pm}$ 
at the scale $\mu =1/a$ are given in Table~\ref{tab:susybp}. 
We find reasonable magnitude of corrections to the tree-level results
for all the renormalization factors.

%
\section{Conclusion}
\label{sec:conclusion} 
%
In this paper we have evaluated the renormalization factors 
for  the generic four-quark operators
at the one-loop level
in DWQCD with the plaquette and Iwasaki gauge actions.
As an application we explain how to construct the renormalization
factors for the SUSY operators by using our results.
We also show that taking the parameters employed in Ref.\cite{Noaki03}
the numerical values for the 
renormalization factors result in
reasonable magnitude with the mean-field improvement.
%
%
\section*{Acknowledgments}
%
This work is supported in part by Grants-in-Aid of the Ministory
of Education (No.15740165).
%
%

%


%
\newcommand{\J}[4]{{#1} {\bf #2} (#3) #4}
\newcommand{\AP}{Ann.~Phys.}
\newcommand{\CMP}{Commun.~Math.~Phys.}
\newcommand{\EUR}{Eur.~Phys.J}
\newcommand{\IJMP}{Int.~J.~Mod.~Phys.}
\newcommand{\MPL}{Mod.~Phys.~Lett.}
\newcommand{\NP}{Nucl.~Phys.}
\newcommand{\NPSup}{Nucl.~Phys.~B (Proc.~Suppl.)}
\newcommand{\PL}{Phys.~Lett.}
\newcommand{\PR}{Phys.~Rev.}
\newcommand{\PRL}{Phys.~Rev.~Lett.}
\newcommand{\PTP}{Prog. Theor. Phys.}
\newcommand{\Suppl}{Prog. Theor. Phys. Suppl.}

\newpage

\newpage
\begin{table}[t]
\begin{center}
\caption{Numerical values for $v_{ij}$ $(i,j=1,\dots,5)$ as a
function of $M$ with plaquette gauge action.}
\begin{tabular}{|cccccc|}
\hline
\hline
 $M$ \hspace{0.8cm} & \hspace{0.8cm}$v_{11}^+$\hspace{0.8cm} & \hspace{0.8cm}$v_{11}^-$\hspace{0.8cm} & \hspace{0.8cm}$v_{22}^\pm$\hspace{0.8cm} & \hspace{0.8cm}$v_{23}^+$\hspace{0.8cm} & \hspace{0.8cm}$v_{23}^-$\hspace{0.8cm}  \\
\hline
$0.1$ & $13.5696(73)$&$11.5369(73)$&$13.2308(18)$&$2.0327(66)$&$-2.0327(66)$ \\ 
$0.2$ & $13.0548(73)$&$12.5870(73)$&$12.9768(18)$&$0.4678(66)$&$-0.4678(66)$ \\ 
$0.3$ & $12.6404(73)$&$13.4380(73)$&$12.7733(18)$&$-0.7976(66)$&$0.7976(66)$ \\ 
$0.4$ & $12.2775(72)$&$14.1882(72)$&$12.5960(18)$&$-1.9107(65)$&$1.9107(65)$ \\ 
$0.5$ & $11.9450(73)$&$14.8801(73)$&$12.4342(18)$&$-2.9352(66)$&$2.9352(66)$ \\ 
$0.6$ & $11.6307(72)$&$15.5380(72)$&$12.2819(18)$&$-3.9073(65)$&$3.9073(65)$ \\ 
$0.7$ & $11.3269(73)$&$16.1781(73)$&$12.1354(18)$&$-4.8512(66)$&$4.8512(66)$ \\ 
$0.8$ & $11.0275(72)$&$16.8129(72)$&$11.9918(18)$&$-5.7853(65)$&$5.7853(65)$ \\ 
$0.9$ & $10.7276(74)$&$17.4523(74)$&$11.8484(18)$&$-6.7247(66)$&$6.7247(66)$ \\ 
$1.0$ & $10.4225(73)$&$18.1071(73)$&$11.7033(18)$&$-7.6845(65)$&$7.6845(65)$ \\ 
$1.1$ & $10.1073(73)$&$18.7869(73)$&$11.5539(18)$&$-8.6796(66)$&$8.6796(66)$ \\ 
$1.2$ & $9.7767(75)$&$19.5039(75)$&$11.3979(19)$&$-9.7271(67)$&$9.7271(67)$ \\ 
$1.3$ & $9.4243(73)$&$20.2719(73)$&$11.2322(18)$&$-10.8475(66)$&$10.8475(66)$ \\ 
$1.4$ & $9.0419(75)$&$21.1082(75)$&$11.0529(19)$&$-12.0664(67)$&$12.0664(67)$ \\ 
$1.5$ & $8.6183(73)$&$22.0373(73)$&$10.8548(18)$&$-13.4191(65)$&$13.4191(65)$ \\ 
$1.6$ & $8.1375(74)$&$23.0928(74)$&$10.6301(18)$&$-14.9553(66)$&$14.9553(66)$ \\ 
$1.7$ & $7.5747(73)$&$24.3277(73)$&$10.3669(18)$&$-16.7531(66)$&$16.7531(66)$ \\ 
$1.8$ & $6.8864(73)$&$25.8324(73)$&$10.0440(18)$&$-18.9460(66)$&$18.9460(66)$ \\ 
$1.9$ & $5.9812(73)$&$27.7944(73)$&$9.6168(18)$&$-21.8132(66)$&$21.8132(66)$ \\ 
\hline
\hline
\end{tabular}
\end{center}
\end{table}

\begin{table}[b]
\begin{center}
\caption{Numerical values for $v_{ij}$ $(i,j=1,\dots,5)$ as a
function of $M$ with plaquette gauge action.}
\begin{tabular}{|llllll|}
\hline
\hline
 $M$\hspace{0.8cm} &\hspace{0.8cm}$v_{33}^\pm$\hspace{0.8cm} &\hspace{0.8cm}$v_{44}^+$
 \hspace{0.8cm} & \hspace{0.8cm}$v_{44}^-$\hspace{0.8cm} & \hspace{0.8cm}$v_{45}^+$
 \hspace{0.8cm} &\hspace{0.8cm}$v_{45}^-$\hspace{0.8cm}\\
\hline
$0.1$ & $10.182(12)$&$11.1981(84)$&$9.165(15)$&$0.11293(37)$&$-0.5646(18)$ \\ 
$0.2$ & $12.275(12)$&$12.5090(84)$&$12.041(15)$&$0.02599(37)$&$-0.1299(18)$ \\ 
$0.3$ & $13.970(12)$&$13.5709(84)$&$14.369(15)$&$-0.04431(37)$&$0.2216(18)$ \\ 
$0.4$ & $15.462(12)$&$14.5067(83)$&$16.417(15)$&$-0.10615(36)$&$0.5307(18)$ \\ 
$0.5$ & $16.837(12)$&$15.3693(84)$&$18.305(15)$&$-0.16307(36)$&$0.8153(18)$ \\ 
$0.6$ & $18.143(12)$&$16.1893(83)$&$20.097(15)$&$-0.21707(36)$&$1.0854(18)$ \\ 
$0.7$ & $19.412(12)$&$16.9866(84)$&$21.838(15)$&$-0.26951(37)$&$1.3476(18)$ \\ 
$0.8$ & $20.670(11)$&$17.7771(83)$&$23.562(15)$&$-0.32141(36)$&$1.6070(18)$ \\ 
$0.9$ & $21.935(12)$&$18.5731(85)$&$25.298(15)$&$-0.37359(37)$&$1.8680(18)$ \\ 
$1.0$ & $23.230(12)$&$19.3878(84)$&$27.072(15)$&$-0.42692(36)$&$2.1346(18)$ \\ 
$1.1$ & $24.573(12)$&$20.2335(84)$&$28.913(15)$&$-0.48220(37)$&$2.4110(18)$ \\ 
$1.2$ & $25.989(12)$&$21.1250(86)$&$30.852(15)$&$-0.54040(37)$&$2.7020(19)$ \\ 
$1.3$ & $27.504(12)$&$22.0798(84)$&$32.927(15)$&$-0.60264(37)$&$3.0132(18)$ \\ 
$1.4$ & $29.153(12)$&$23.1193(86)$&$35.186(15)$&$-0.67036(37)$&$3.3518(19)$ \\ 
$1.5$ & $30.983(12)$&$24.2739(84)$&$37.693(15)$&$-0.74550(36)$&$3.7275(18)$ \\ 
$1.6$ & $33.063(12)$&$25.5853(85)$&$40.541(15)$&$-0.83085(37)$&$4.1542(18)$ \\ 
$1.7$ & $35.496(12)$&$27.1199(84)$&$43.873(15)$&$-0.93073(37)$&$4.6536(18)$ \\ 
$1.8$ & $38.463(12)$&$28.9900(84)$&$47.936(15)$&$-1.05255(37)$&$5.2628(18)$ \\ 
$1.9$ & $42.337(12)$&$31.4299(84)$&$53.243(15)$&$-1.21184(37)$&$6.0592(18)$ \\ 
\hline
\hline
\end{tabular}
\end{center}
\end{table}

\clearpage
\newpage
\begin{table}[t]
\begin{center}
\caption{Numerical values for $v_{ij}$ $(i,j=1,\dots,5)$ as a
function of $M$ with plaquette gauge action.}
\begin{tabular}{|ccccc|}
\hline
\hline
  $M$\hspace{0.8cm} &\hspace{0.8cm}$v_{55}^+$\hspace{0.8cm} &\hspace{0.8cm}$v_{55}^-$
 \hspace{0.8cm} & \hspace{0.8cm}$v_{54}^+$\hspace{0.8cm} & \hspace{0.8cm}$v_{54}^-$
 \hspace{0.8cm}\\
\hline
$0.1$ & $14.8119(33)$&$12.7791(33)$&$-1.6939(55)$&$0.3388(11)$ \\ 
$0.2$ & $13.3406(33)$&$12.8729(33)$&$-0.3898(55)$&$0.0780(11)$ \\ 
$0.3$ & $12.1530(33)$&$12.9506(33)$&$0.6647(55)$&$-0.1329(11)$ \\ 
$0.4$ & $11.1099(32)$&$13.0206(33)$&$1.5922(54)$&$-0.3184(11)$ \\ 
$0.5$ & $10.1512(33)$&$13.0864(33)$&$2.4460(55)$&$-0.4892(11)$ \\ 
$0.6$ & $9.2429(33)$&$13.1502(33)$&$3.2561(54)$&$-0.6512(11)$ \\ 
$0.7$ & $8.3622(33)$&$13.2135(33)$&$4.0427(55)$&$-0.8085(11)$ \\ 
$0.8$ & $7.4920(32)$&$13.2774(32)$&$4.8211(54)$&$-0.9642(11)$ \\ 
$0.9$ & $6.6181(33)$&$13.3428(33)$&$5.6039(55)$&$-1.1208(11)$ \\ 
$1.0$ & $5.7264(33)$&$13.4109(33)$&$6.4038(55)$&$-1.2808(11)$ \\ 
$1.1$ & $4.8031(33)$&$13.4827(33)$&$7.2330(55)$&$-1.4466(11)$ \\ 
$1.2$ & $3.8324(34)$&$13.5595(34)$&$8.1059(56)$&$-1.6212(11)$ \\ 
$1.3$ & $2.7953(33)$&$13.6428(33)$&$9.0396(55)$&$-1.8079(11)$ \\ 
$1.4$ & $1.6679(34)$&$13.7343(34)$&$10.0553(56)$&$-2.0111(11)$ \\ 
$1.5$ & $0.4177(33)$&$13.8368(33)$&$11.1826(55)$&$-2.2365(11)$ \\ 
$1.6$ & $-1.0018(33)$&$13.9535(33)$&$12.4627(55)$&$-2.4925(11)$ \\ 
$1.7$ & $-2.6633(33)$&$14.0898(33)$&$13.9609(55)$&$-2.7922(11)$ \\ 
$1.8$ & $-4.6917(33)$&$14.2543(33)$&$15.7883(55)$&$-3.1577(11)$ \\ 
$1.9$ & $-7.3491(33)$&$14.4641(33)$&$18.1777(55)$&$-3.6355(11)$ \\ 
\hline
\hline
\end{tabular}
\end{center}
\end{table}

\begin{table}[b]
\begin{center}
\caption{Numerical values for $z_{ij}$ $(i,j=1,\dots,5)$ as a
function of $M$ with plaquette gauge action.}
\begin{tabular}{|llllll|}
\hline
\hline
$M$\hspace{0.8cm} &\hspace{0.8cm}$z_{11}^+$\hspace{0.8cm} &\hspace{0.8cm}$z_{11}^-$
 \hspace{0.8cm} &\hspace{0.8cm}$z_{22}^\pm$\hspace{0.8cm} &
  \hspace{0.8cm}$z_{23}^+$\hspace{0.8cm} &
\hspace{0.8cm}$z_{23}^-$\hspace{0.8cm}\\
\hline
$0.1$ & $-52.33012(24)$&$-39.2974(68)$&$-48.1580(13)$&$-3.0327(66)$&$1.0327(66)$ \\ 
$0.2$ & $-51.4142(13)$&$-39.9464(53)$&$-47.50286(22)$&$-1.4678(66)$&$-0.5322(66)$ \\ 
$0.3$ & $-50.6609(11)$&$-40.4585(55)$&$-46.960537(34)$&$-0.2024(66)$&$-1.7976(66)$ \\ 
$0.4$ & $-50.0050(14)$&$-40.9157(51)$&$-46.49015(35)$&$0.9107(65)$&$-2.9107(65)$ \\ 
$0.5$ & $-49.41704(64)$&$-41.3522(72)$&$-46.0729(17)$&$1.9352(66)$&$-3.9352(66)$ \\ 
$0.6$ & $-48.8806(11)$&$-41.7879(54)$&$-45.698460(58)$&$2.9073(65)$&$-4.9073(65)$ \\ 
$0.7$ & $-48.3854(18)$&$-42.2366(48)$&$-45.36057(67)$&$3.8512(66)$&$-5.8512(66)$ \\ 
$0.8$ & $-47.92464(38)$&$-42.7100(61)$&$-45.05553(70)$&$4.7853(65)$&$-6.7853(65)$ \\ 
$0.9$ & $-47.49355(14)$&$-43.2182(68)$&$-44.7810(12)$&$5.7247(66)$&$-7.7247(66)$ \\ 
$1.0$ & $-47.0888(16)$&$-43.7734(50)$&$-44.53624(49)$&$6.6845(65)$&$-8.6845(65)$ \\ 
$1.1$ & $-46.707975(79)$&$-44.3876(67)$&$-44.3212(12)$&$7.6796(66)$&$-9.6796(66)$ \\ 
$1.2$ & $-46.3494(12)$&$-45.0765(55)$&$-44.137215(96)$&$8.7271(67)$&$-10.7271(67)$ \\ 
$1.3$ & $-46.011799(83)$&$-45.8593(65)$&$-43.9864(10)$&$9.8475(66)$&$-11.8475(66)$ \\ 
$1.4$ & $-45.6944(11)$&$-46.7608(56)$&$-43.872137(23)$&$11.0664(67)$&$-13.0664(67)$ \\ 
$1.5$ & $-45.3964(14)$&$-47.8155(51)$&$-43.79961(34)$&$12.4191(65)$&$-14.4191(65)$ \\ 
$1.6$ & $-45.11658(35)$&$-49.0719(70)$&$-43.7758(15)$&$13.9553(66)$&$-15.9553(66)$ \\ 
$1.7$ & $-44.8518(17)$&$-50.6049(49)$&$-43.81064(62)$&$15.7531(66)$&$-17.7531(66)$ \\ 
$1.8$ & $-44.59306(72)$&$-52.5390(73)$&$-43.9174(18)$&$17.9460(66)$&$-19.9460(66)$ \\ 
$1.9$ & $-44.3086(13)$&$-55.1218(53)$&$-44.11077(18)$&$20.8132(66)$&$-22.8132(66)$ \\ 
\hline
\hline
\end{tabular}
\end{center}
\end{table}

\clearpage
\newpage
\begin{table}[t]
\begin{center}
\caption{Numerical values for $z_{ij}$ $(i,j=1,\dots,5)$ as a
function of $M$ with plaquette gauge action.}
\begin{tabular}{|llllll|}
\hline
\hline
  $M$\hspace{0.8cm} &\hspace{0.8cm}$z_{33}^\pm$\hspace{0.8cm} &\hspace{0.8cm}$z_{44}^+$
 \hspace{0.8cm} & \hspace{0.8cm}$z_{44}^-$\hspace{0.8cm} & \hspace{0.8cm}$z_{45}^+$
 \hspace{0.8cm} &\hspace{0.8cm}$z_{45}^-$\hspace{0.8cm}\\
\hline
$0.1$ & $-37.609(11)$&$-40.2919(79)$&$-36.259(14)$&$-0.77960(37)$&$1.8980(18)$ \\ 
$0.2$ & $-39.3012(97)$&$-41.2018(64)$&$-38.734(13)$&$-0.69265(37)$&$1.4633(18)$ \\ 
$0.3$ & $-40.6569(99)$&$-41.9248(66)$&$-40.722(13)$&$-0.62236(37)$&$1.1118(18)$ \\ 
$0.4$ & $-41.8562(94)$&$-42.5675(61)$&$-42.478(13)$&$-0.56052(36)$&$0.8026(18)$ \\ 
$0.5$ & $-42.976(12)$&$-43.1748(83)$&$-44.110(15)$&$-0.50360(36)$&$0.5180(18)$ \\ 
$0.6$ & $-44.0595(97)$&$-43.7725(65)$&$-45.680(13)$&$-0.44959(36)$&$0.2480(18)$ \\ 
$0.7$ & $-45.1374(92)$&$-44.3785(59)$&$-47.230(13)$&$-0.39715(37)$&$-0.0142(18)$ \\ 
$0.8$ & $-46.234(10)$&$-45.0075(72)$&$-48.793(14)$&$-0.34526(36)$&$-0.2737(18)$ \\ 
$0.9$ & $-47.368(11)$&$-45.6723(79)$&$-50.397(14)$&$-0.29307(37)$&$-0.5346(18)$ \\ 
$1.0$ & $-48.5631(93)$&$-46.3875(61)$&$-52.072(13)$&$-0.23975(36)$&$-0.8013(18)$ \\ 
$1.1$ & $-49.841(11)$&$-47.1675(78)$&$-53.847(14)$&$-0.18447(37)$&$-1.0777(18)$ \\ 
$1.2$ & $-51.228(10)$&$-48.0310(66)$&$-55.758(13)$&$-0.12627(37)$&$-1.3686(19)$ \\ 
$1.3$ & $-52.758(11)$&$-49.0006(76)$&$-57.848(14)$&$-0.06403(37)$&$-1.6799(18)$ \\ 
$1.4$ & $-54.472(10)$&$-50.1052(67)$&$-60.172(13)$&$0.00369(37)$&$-2.0184(19)$ \\ 
$1.5$ & $-56.4282(95)$&$-51.3854(62)$&$-62.804(13)$&$0.07884(36)$&$-2.3942(18)$ \\ 
$1.6$ & $-58.709(11)$&$-52.8977(81)$&$-65.853(15)$&$0.16418(37)$&$-2.8209(18)$ \\ 
$1.7$ & $-61.4402(93)$&$-54.7304(60)$&$-69.483(13)$&$0.26406(37)$&$-3.3203(18)$ \\ 
$1.8$ & $-64.836(12)$&$-57.0300(84)$&$-73.976(15)$&$0.38589(37)$&$-3.9294(18)$ \\ 
$1.9$ & $-69.3305(97)$&$-60.0906(64)$&$-79.904(13)$&$0.54518(37)$&$-4.7259(18)$ \\ 
\hline
\hline
\end{tabular}
\end{center}
\end{table}

\begin{table}[b]
\begin{center}
\caption{Numerical values for $z_{ij}$ $(i,j=1,\dots,5)$ as a
function of $M$ with plaquette gauge action.}
\begin{tabular}{|ccccc|}
\hline
\hline
  $M$\hspace{0.8cm} &\hspace{0.8cm}$z_{55}^+$\hspace{0.8cm} &
  \hspace{0.8cm}$z_{55}^-$\hspace{0.8cm} &
  \hspace{0.8cm}$z_{54}^+$\hspace{0.8cm} &
\hspace{0.8cm}$z_{54}^-$\hspace{0.8cm}\\
\hline  
$0.1$ & $-50.2390(38)$&$-44.2063(28)$&$2.6939(55)$&$0.6612(11)$ \\ 
$0.2$ & $-48.3667(54)$&$-43.8989(12)$&$1.3898(55)$&$0.9220(11)$ \\ 
$0.3$ & $-46.8402(52)$&$-43.6378(14)$&$0.3353(55)$&$1.1329(11)$ \\ 
$0.4$ & $-45.5041(54)$&$-43.4147(11)$&$-0.5922(54)$&$1.3184(11)$ \\ 
$0.5$ & $-44.2900(34)$&$-43.2252(32)$&$-1.4460(55)$&$1.4892(11)$ \\ 
$0.6$ & $-43.1594(51)$&$-43.0668(14)$&$-2.2561(54)$&$1.6512(11)$ \\ 
$0.7$ & $-42.0874(58)$&$-42.93862(80)$&$-3.0427(55)$&$1.8085(11)$ \\ 
$0.8$ & $-41.0558(43)$&$-42.8412(21)$&$-3.8211(54)$&$1.9642(11)$ \\ 
$0.9$ & $-40.0507(39)$&$-42.7754(27)$&$-4.6039(55)$&$2.1208(11)$ \\ 
$1.0$ & $-39.0594(56)$&$-42.74392(96)$&$-5.4038(55)$&$2.2808(11)$ \\ 
$1.1$ & $-38.0704(40)$&$-42.7500(26)$&$-6.2330(55)$&$2.4466(11)$ \\ 
$1.2$ & $-37.0717(53)$&$-42.7988(14)$&$-7.1059(56)$&$2.6212(11)$ \\ 
$1.3$ & $-36.0494(41)$&$-42.8970(25)$&$-8.0396(55)$&$2.8079(11)$ \\ 
$1.4$ & $-34.9872(52)$&$-43.0536(15)$&$-9.0553(56)$&$3.0111(11)$ \\ 
$1.5$ & $-33.8626(54)$&$-43.2816(11)$&$-10.1826(55)$&$3.2365(11)$ \\ 
$1.6$ & $-32.6439(37)$&$-43.5992(29)$&$-11.4627(55)$&$3.4925(11)$ \\ 
$1.7$ & $-31.2805(57)$&$-44.03354(85)$&$-12.9609(55)$&$3.7922(11)$ \\ 
$1.8$ & $-29.6816(33)$&$-44.6276(33)$&$-14.7883(55)$&$4.1577(11)$ \\ 
$1.9$ & $-27.6450(53)$&$-45.4581(13)$&$-17.1777(55)$&$4.6355(11)$ \\
\hline
\hline
\end{tabular}
\end{center}
\end{table}

\clearpage
\newpage

\begin{table}[t]
\begin{center}
\caption{Numerical values for$v_{ij}$ $(i,j=1,\dots,5)$ as a function 
of $M$ with Iwasaki gauge action.}
\begin{tabular}{|llllll|}
\hline
\hline
  $M$\hspace{0.8cm} &\hspace{0.8cm}$v_{11}^+$\hspace{0.8cm} &\hspace{0.8cm}$v_{11}^-$
 \hspace{0.8cm} &\hspace{0.8cm}$v_{22}^\pm$\hspace{0.8cm} &\hspace{0.8cm}$v_{23}^+$
 \hspace{0.8cm}&\hspace{0.8cm}$v_{23}^-$\hspace{0.8cm}\\
\hline
$0.1$ & $14.1911(73)$&$10.1069(73)$&$13.5104(18)$&$4.0842(66)$&$-4.0842(66)$ \\ 
$0.2$ & $13.7193(73)$&$11.0602(73)$&$13.2761(18)$&$2.6591(66)$&$-2.6591(66)$ \\ 
$0.3$ & $13.3494(73)$&$11.8105(73)$&$13.0929(18)$&$1.5389(66)$&$-1.5389(66)$ \\ 
$0.4$ & $13.0327(72)$&$12.4558(72)$&$12.9365(18)$&$0.5769(65)$&$-0.5769(65)$ \\ 
$0.5$ & $12.7480(73)$&$13.0378(73)$&$12.7963(18)$&$-0.2899(66)$&$0.2899(66)$ \\ 
$0.6$ & $12.4836(72)$&$13.5804(72)$&$12.6664(18)$&$-1.0968(65)$&$1.0968(65)$ \\ 
$0.7$ & $12.2320(73)$&$14.0990(73)$&$12.5431(18)$&$-1.8670(66)$&$1.8670(66)$ \\ 
$0.8$ & $11.9874(72)$&$14.6053(72)$&$12.4237(18)$&$-2.6179(65)$&$2.6179(65)$ \\ 
$0.9$ & $11.7451(74)$&$15.1082(74)$&$12.3057(18)$&$-3.3631(66)$&$3.3631(66)$ \\ 
$1.0$ & $11.5011(73)$&$15.6173(73)$&$12.1871(18)$&$-4.1162(65)$&$4.1162(65)$ \\ 
$1.1$ & $11.2508(73)$&$16.1409(73)$&$12.0658(18)$&$-4.8901(66)$&$4.8901(66)$ \\ 
$1.2$ & $10.9896(75)$&$16.6892(75)$&$11.9395(19)$&$-5.6997(67)$&$5.6997(67)$ \\ 
$1.3$ & $10.7118(73)$&$17.2742(73)$&$11.8055(18)$&$-6.5624(66)$&$6.5624(66)$ \\ 
$1.4$ & $10.4101(75)$&$17.9105(75)$&$11.6602(19)$&$-7.5004(67)$&$7.5004(67)$ \\ 
$1.5$ & $10.0748(73)$&$18.6191(73)$&$11.4988(18)$&$-8.5444(65)$&$8.5444(65)$ \\ 
$1.6$ & $9.6912(74)$&$19.4294(74)$&$11.3142(18)$&$-9.7382(66)$&$9.7382(66)$ \\ 
$1.7$ & $9.2365(73)$&$20.3887(73)$&$11.0952(18)$&$-11.1523(66)$&$11.1523(66)$ \\ 
$1.8$ & $8.6700(73)$&$21.5799(73)$&$10.8217(18)$&$-12.9099(66)$&$12.9099(66)$ \\ 
$1.9$ & $7.9044(73)$&$23.1801(73)$&$10.4503(18)$&$-15.2758(66)$&$15.2758(66)$ \\
\hline
\hline
\end{tabular}
\end{center}
\end{table}

\begin{table}[b]
\begin{center}
\caption{Numerical values for $v_{ij}$ $(i,j=1,\dots,5)$ 
as a function of $M$ with Iwasaki gauge action.}
\begin{tabular}{|llllll|}
\hline
\hline
  $M$\hspace{0.8cm} &\hspace{0.8cm}$v_{33}^\pm$\hspace{0.8cm} &\hspace{0.8cm}$v_{44}^+$
 \hspace{0.8cm} &\hspace{0.8cm}$v_{44}^-$\hspace{0.8cm} &\hspace{0.8cm}$v_{45}^+$
 \hspace{0.8cm} &\hspace{0.8cm}$v_{45}^-$\hspace{0.8cm} \\
\hline
$0.1$ & $7.384(12)$&$9.4262(84)$&$5.342(15)$&$0.22690(37)$&$-1.1345(18)$ \\ 
$0.2$ & $9.287(12)$&$10.6170(84)$&$7.958(15)$&$0.14773(37)$&$-0.7386(18)$ \\ 
$0.3$ & $10.785(12)$&$11.5540(84)$&$10.015(15)$&$0.08549(37)$&$-0.4275(18)$ \\ 
$0.4$ & $12.071(12)$&$12.3596(83)$&$11.783(15)$&$0.03205(36)$&$-0.1603(18)$ \\ 
$0.5$ & $13.231(12)$&$13.0862(84)$&$13.376(15)$&$-0.01610(36)$&$0.0805(18)$ \\ 
$0.6$ & $14.312(12)$&$13.7632(83)$&$14.860(15)$&$-0.06093(36)$&$0.3047(18)$ \\ 
$0.7$ & $15.344(12)$&$14.4102(84)$&$16.277(15)$&$-0.10372(37)$&$0.5186(18)$ \\ 
$0.8$ & $16.351(11)$&$15.0416(83)$&$17.660(15)$&$-0.14544(36)$&$0.7272(18)$ \\ 
$0.9$ & $17.350(12)$&$15.6687(85)$&$19.032(15)$&$-0.18684(37)$&$0.9342(18)$ \\ 
$1.0$ & $18.361(12)$&$16.3033(84)$&$20.420(15)$&$-0.22868(36)$&$1.1434(18)$ \\ 
$1.1$ & $19.401(12)$&$16.9559(84)$&$21.846(15)$&$-0.27167(37)$&$1.3584(18)$ \\ 
$1.2$ & $20.489(12)$&$17.6392(86)$&$23.339(15)$&$-0.31665(37)$&$1.5832(19)$ \\ 
$1.3$ & $21.649(12)$&$18.3680(84)$&$24.930(15)$&$-0.36458(37)$&$1.8229(18)$ \\ 
$1.4$ & $22.911(12)$&$19.1606(86)$&$26.661(15)$&$-0.41669(37)$&$2.0834(19)$ \\ 
$1.5$ & $24.315(12)$&$20.0432(84)$&$28.588(15)$&$-0.47469(36)$&$2.3734(18)$ \\ 
$1.6$ & $25.921(12)$&$21.0524(85)$&$30.791(15)$&$-0.54101(37)$&$2.7051(18)$ \\ 
$1.7$ & $27.824(12)$&$22.2475(84)$&$33.400(15)$&$-0.61957(37)$&$3.0979(18)$ \\ 
$1.8$ & $30.186(12)$&$23.7315(84)$&$36.641(15)$&$-0.71721(37)$&$3.5861(18)$ \\ 
$1.9$ & $33.364(12)$&$25.7261(84)$&$41.002(15)$&$-0.84865(37)$&$4.2433(18)$ \\ 
\hline
\hline
\end{tabular}
\end{center}
\end{table}

\clearpage
\newpage
\begin{table}[t]
\begin{center}
\caption{Numerical values for
$v_{ij}$ $(i,j=1,\dots,5)$ as a function of $M$ with Iwasaki gauge action.}
\begin{tabular}{|lllll|}
\hline
\hline
  $M$\hspace{0.8cm} &\hspace{0.8cm}$v_{55}^+$\hspace{0.8cm} &\hspace{0.8cm}$v_{55}^-$
 \hspace{0.8cm} &\hspace{0.8cm}$v_{54}^+$\hspace{0.8cm} &\hspace{0.8cm}$v_{54}^-$
 \hspace{0.8cm}\\
\hline
$0.1$ & $16.6870(33)$&$12.6028(33)$&$-3.4035(55)$&$0.6807(11)$ \\ 
$0.2$ & $15.3443(33)$&$12.6852(33)$&$-2.2159(55)$&$0.4432(11)$ \\ 
$0.3$ & $14.2899(33)$&$12.7510(33)$&$-1.2824(55)$&$0.2565(11)$ \\ 
$0.4$ & $13.3852(32)$&$12.8083(33)$&$-0.4808(54)$&$0.0962(11)$ \\ 
$0.5$ & $12.5708(33)$&$12.8607(33)$&$0.2416(55)$&$-0.0483(11)$ \\ 
$0.6$ & $11.8133(33)$&$12.9102(33)$&$0.9140(54)$&$-0.1828(11)$ \\ 
$0.7$ & $11.0910(33)$&$12.9580(33)$&$1.5559(55)$&$-0.3112(11)$ \\ 
$0.8$ & $10.3875(32)$&$13.0055(32)$&$2.1816(54)$&$-0.4363(11)$ \\ 
$0.9$ & $9.6899(33)$&$13.0530(33)$&$2.8025(55)$&$-0.5605(11)$ \\ 
$1.0$ & $8.9856(33)$&$13.1018(33)$&$3.4302(55)$&$-0.6860(11)$ \\ 
$1.1$ & $8.2624(33)$&$13.1525(33)$&$4.0751(55)$&$-0.8150(11)$ \\ 
$1.2$ & $7.5064(34)$&$13.2061(34)$&$4.7497(56)$&$-0.9499(11)$ \\ 
$1.3$ & $6.7014(33)$&$13.2638(33)$&$5.4687(55)$&$-1.0937(11)$ \\ 
$1.4$ & $5.8266(34)$&$13.3270(34)$&$6.2503(56)$&$-1.2501(11)$ \\ 
$1.5$ & $4.8532(33)$&$13.3976(33)$&$7.1203(55)$&$-1.4241(11)$ \\ 
$1.6$ & $3.7400(33)$&$13.4782(33)$&$8.1152(55)$&$-1.6230(11)$ \\ 
$1.7$ & $2.4212(33)$&$13.5735(33)$&$9.2936(55)$&$-1.8587(11)$ \\ 
$1.8$ & $0.7807(33)$&$13.6905(33)$&$10.7582(55)$&$-2.1516(11)$ \\ 
$1.9$ & $-1.4308(33)$&$13.8449(33)$&$12.7298(55)$&$-2.5460(11)$ \\ 
\hline
\hline
\end{tabular}
\end{center}
\end{table}

\begin{table}[b]
\begin{center}
\caption{Numerical values for
$z_{ij}$ $(i,j=1,\dots,5)$ as a function of $M$ with Iwasaki gauge action.}
\begin{tabular}{|llllll|}
\hline
\hline
  $M$\hspace{0.8cm} &\hspace{0.8cm}$z_{11}^+$\hspace{0.8cm} &\hspace{0.8cm}$z_{11}^-$
 \hspace{0.8cm} &\hspace{0.8cm}$z_{22}^\pm$\hspace{0.8cm} &\hspace{0.8cm}$z_{23}^+$
 \hspace{0.8cm} &\hspace{0.8cm}$z_{23}^-$\hspace{0.8cm}\\
\hline
$0.1$ & $-30.25980(58)$&$-15.1756(72)$&$-25.7458(17)$&$-5.0842(66)$&$3.0842(66)$ \\ 
$0.2$ & $-29.43439(58)$&$-15.7753(72)$&$-25.1579(17)$&$-3.6591(66)$&$1.6591(66)$ \\ 
$0.3$ & $-28.77133(70)$&$-16.2324(73)$&$-24.6815(18)$&$-2.5389(66)$&$0.5389(66)$ \\ 
$0.4$ & $-28.20508(57)$&$-16.6282(71)$&$-24.2756(17)$&$-1.5769(65)$&$-0.4231(65)$ \\ 
$0.5$ & $-27.70614(68)$&$-16.9960(72)$&$-23.9211(18)$&$-0.7101(66)$&$-1.2899(66)$ \\ 
$0.6$ & $-27.25798(48)$&$-17.3548(70)$&$-23.6075(16)$&$0.0968(65)$&$-2.0968(65)$ \\ 
$0.7$ & $-26.85022(60)$&$-17.7173(72)$&$-23.3281(17)$&$0.8670(66)$&$-2.8670(66)$ \\ 
$0.8$ & $-26.47585(51)$&$-18.0938(70)$&$-23.0788(16)$&$1.6179(65)$&$-3.6179(65)$ \\ 
$0.9$ & $-26.12985(57)$&$-18.4929(72)$&$-22.8570(17)$&$2.3631(66)$&$-4.3631(66)$ \\ 
$1.0$ & $-25.80864(62)$&$-18.9249(72)$&$-22.6613(17)$&$3.1162(65)$&$-5.1162(65)$ \\ 
$1.1$ & $-25.50947(72)$&$-19.3996(73)$&$-22.4912(18)$&$3.8901(66)$&$-5.8901(66)$ \\ 
$1.2$ & $-25.23026(60)$&$-19.9299(73)$&$-22.3469(17)$&$4.6997(67)$&$-6.6997(67)$ \\ 
$1.3$ & $-24.96938(47)$&$-20.5318(71)$&$-22.2298(16)$&$5.5624(66)$&$-7.5624(66)$ \\ 
$1.4$ & $-24.72536(72)$&$-21.2258(74)$&$-22.1421(18)$&$6.5004(67)$&$-8.5004(67)$ \\ 
$1.5$ & $-24.49673(63)$&$-22.0411(72)$&$-22.0875(17)$&$7.5444(65)$&$-9.5444(65)$ \\ 
$1.6$ & $-24.28127(49)$&$-23.0195(71)$&$-22.0710(16)$&$8.7382(66)$&$-10.7382(66)$ \\ 
$1.7$ & $-24.07481(51)$&$-24.2271(71)$&$-22.1002(16)$&$10.1523(66)$&$-12.1523(66)$ \\ 
$1.8$ & $-23.86686(71)$&$-25.7767(73)$&$-22.1852(18)$&$11.9099(66)$&$-13.9099(66)$ \\ 
$1.9$ & $-23.62377(52)$&$-27.8995(71)$&$-22.3364(16)$&$14.2758(66)$&$-16.2758(66)$ \\ 
\hline
\hline
\end{tabular}
\end{center}
\end{table}

\clearpage
\newpage
\begin{table}[t]
\begin{center}
\caption{Numerical values for
$z_{ij}$ $(i,j=1,\dots,5)$ as a function of $M$ with Iwasaki gauge action.}
\begin{tabular}{|llllll|}
\hline
\hline
  $M$\hspace{0.8cm} &\hspace{0.8cm}$z_{33}^\pm$\hspace{0.8cm} &\hspace{0.8cm}$z_{44}^+$
 \hspace{0.8cm} &\hspace{0.8cm}$z_{44}^-$\hspace{0.8cm} &\hspace{0.8cm}$z_{45}^+$
 \hspace{0.8cm}&\hspace{0.8cm}$z_{45}^-$\hspace{0.8cm}\\
\hline
$0.1$ & $-12.119(12)$&$-15.8282(82)$&$-9.744(15)$ &$-0.89357(37)$&$2.4678(18)$ \\ 
$0.2$ & $-13.669(12)$&$-16.6654(83)$&$-12.006(15)$&$-0.81440(37)$&$2.0720(18)$ \\ 
$0.3$ & $-14.873(12)$&$-17.3093(84)$&$-13.770(15)$&$-0.75216(37)$&$1.7608(18)$ \\ 
$0.4$ & $-15.910(11)$&$-17.8654(82)$&$-15.288(15)$&$-0.69872(36)$&$1.4936(18)$ \\ 
$0.5$ & $-16.856(12)$&$-18.3777(83)$&$-16.668(15)$&$-0.65056(36)$&$1.2528(18)$ \\ 
$0.6$ & $-17.753(11)$&$-18.8709(81)$&$-17.968(15)$&$-0.60573(36)$&$1.0287(18)$ \\ 
$0.7$ & $-18.629(12)$&$-19.3618(83)$&$-19.229(15)$&$-0.56294(37)$&$0.8147(18)$ \\ 
$0.8$ & $-19.506(11)$&$-19.8634(81)$&$-20.481(15)$&$-0.52123(36)$&$0.6061(18)$ \\ 
$0.9$ & $-20.402(12)$&$-20.3867(83)$&$-21.750(15)$&$-0.47983(37)$&$0.3992(18)$ \\ 
$1.0$ & $-21.336(12)$&$-20.9442(83)$&$-23.060(15)$&$-0.43799(36)$&$0.1899(18)$ \\ 
$1.1$ & $-22.326(12)$&$-21.5479(84)$&$-24.438(15)$&$-0.39499(37)$&$-0.0250(18)$ \\ 
$1.2$ & $-23.396(12)$&$-22.2132(84)$&$-25.913(15)$&$-0.35002(37)$&$-0.2499(19)$ \\ 
$1.3$ & $-24.573(11)$&$-22.9589(82)$&$-27.521(15)$&$-0.30209(37)$&$-0.4896(18)$ \\ 
$1.4$ & $-25.893(12)$&$-23.8092(85)$&$-29.310(15)$&$-0.24998(37)$&$-0.7501(19)$ \\ 
$1.5$ & $-27.404(12)$&$-24.7985(83)$&$-31.343(15)$&$-0.19198(36)$&$-1.0401(18)$ \\ 
$1.6$ & $-29.178(12)$&$-25.9758(82)$&$-33.714(15)$&$-0.12566(37)$&$-1.3717(18)$ \\ 
$1.7$ & $-31.329(11)$&$-27.4191(82)$&$-36.571(15)$&$-0.04710(37)$&$-1.7645(18)$ \\ 
$1.8$ & $-34.050(12)$&$-29.2617(84)$&$-40.172(15)$&$0.05055(37)$&$-2.2527(18)$ \\ 
$1.9$ & $-37.750(11)$&$-31.7788(82)$&$-45.055(15)$&$0.18199(37)$&$-2.9099(18)$ \\ 
\hline
\hline
\end{tabular}
\end{center}
\end{table}

\begin{table}[b]
\begin{center}
\caption{Numerical values for
$z_{ij}$ $(i,j=1,\dots,5)$ as a function of $M$ with Iwasaki gauge action.}
\begin{tabular}{|lllll|}
\hline
\hline
  $M$\hspace{0.8cm} &\hspace{0.8cm}$z_{55}^+$\hspace{0.8cm} & \hspace{0.8cm}$z_{55}^-$
 \hspace{0.8cm} & \hspace{0.8cm}$z_{54}^+$\hspace{0.8cm} & \hspace{0.8cm}$z_{54}^-$
 \hspace{0.8cm}\\
\hline
$0.1$ & $-29.4224(34)$&$-21.3382(31)$&$4.4035(55)$&$0.3193(11)$ \\ 
$0.2$ & $-27.7261(34)$&$-21.0670(31)$&$3.2159(55)$&$0.5568(11)$ \\ 
$0.3$ & $-26.3784(33)$&$-20.8395(33)$&$2.2824(55)$&$0.7435(11)$ \\ 
$0.4$ & $-25.2243(34)$&$-20.6474(31)$&$1.4808(54)$&$0.9038(11)$ \\ 
$0.5$ & $-24.1957(33)$&$-20.4855(32)$&$0.7584(55)$&$1.0483(11)$ \\ 
$0.6$ & $-23.2544(35)$&$-20.3512(30)$&$0.0860(54)$&$1.1828(11)$ \\ 
$0.7$ & $-22.3759(34)$&$-20.2430(32)$&$-0.5559(55)$&$1.3112(11)$ \\ 
$0.8$ & $-21.5427(34)$&$-20.1606(30)$&$-1.1816(54)$&$1.4363(11)$ \\ 
$0.9$ & $-20.7413(35)$&$-20.1044(31)$&$-1.8025(55)$&$1.5605(11)$ \\ 
$1.0$ & $-19.9598(34)$&$-20.0761(32)$&$-2.4302(55)$&$1.6860(11)$ \\ 
$1.1$ & $-19.1877(33)$&$-20.0778(33)$&$-3.0751(55)$&$1.8150(11)$ \\ 
$1.2$ & $-18.4138(35)$&$-20.1135(32)$&$-3.7497(56)$&$1.9499(11)$ \\ 
$1.3$ & $-17.6257(36)$&$-20.1881(30)$&$-4.4687(55)$&$2.0937(11)$ \\ 
$1.4$ & $-16.8085(34)$&$-20.3088(33)$&$-5.2503(56)$&$2.2501(11)$ \\ 
$1.5$ & $-15.9418(34)$&$-20.4862(32)$&$-6.1203(55)$&$2.4241(11)$ \\ 
$1.6$ & $-14.9968(36)$&$-20.7350(31)$&$-7.1152(55)$&$2.6230(11)$ \\ 
$1.7$ & $-13.9262(35)$&$-21.0785(31)$&$-8.2936(55)$&$2.8587(11)$ \\ 
$1.8$ & $-12.6442(33)$&$-21.5540(33)$&$-9.7582(55)$&$3.1516(11)$ \\ 
$1.9$ & $-10.9552(35)$&$-22.2310(31)$&$-11.7298(55)$&$3.5460(11)$ \\ 
\hline
\hline
\end{tabular}
\end{center}
\end{table}
\clearpage
\newpage
\begin{table}[t]
\begin{center}
\caption{Numerical values of $Z_{ij}^{\pm}$ $(i,j=1,\dots,5)$
for plaquette gauge action (upper part) 
and Iwasaki gauge action (lower part).}
\label{tab:susybp}
\begin{tabular}{l|llllll|}
\hline
\hline
 $\beta$ &  $P$  & $g_{\msbar}^2(1/a)$ & $M$ &  & \\
 $6.0$ &   $0.59374$ &   $2.1793$ &   $1.80$ &  &\\
\hline
 $\tilde M$& $Z_{11}^+$ &$Z_{11}^-$& $Z_{22}^\pm$ & $Z_{23}^+$
& $Z_{23}^-$ \\
 $1.3112$  &  $0.7287$  &  $0.7289$ &   $0.7564$ & $0.1378$ &   $-0.1654$ \\
&$Z_{33}^+$& $Z_{32}^\pm$  &$Z_{44}^+$& $Z_{44}^-$&  $Z_{45}^+$ \\
  &  $0.6325$ & $\pm0.0207$ &  $0.6853$ &   $0.5613$ &  $-0.0007790$  \\
& $Z_{45}^-$    &$Z_{55}^+$&$Z_{55}^-$& $Z_{54}^+$& $Z_{54}^-$\\
  & $-0.02371$  &$0.8674$  &  $0.7710$ & $-0.1125$ &   $0.03906$\\
\hline
\hline
\hline
 $\beta$ &  $P$  & $R$  &$g_{\msbar}^2(1/a)$ & $M$ & \\
 $2.6$ &  $0.67063$ &   $0.45283$ &   $2.2479$  &  $1.80$ &
 \\
\hline
 $\tilde M$& $Z_{11}^+$ &$Z_{11}^-$& $Z_{22}^\pm$ & $Z_{23}^+$
& $Z_{23}^-$ \\
$1.4198$ &   $0.8062$ &   $0.8531$  &  $0.8425$  &  $0.09548$ &  $-0.1239$\\
  &$Z_{33}^+$& $Z_{32}^\pm$ &$Z_{44}^+$& $Z_{44}^-$&  $Z_{45}^+$\\
 & $0.7847$  & $\pm0.02135$ &$0.8158$ &   $0.7346$ &  $-0.003395$ \\
&$Z_{45}^-$& $Z_{55}^+$&$Z_{55}^-$& $Z_{54}^+$& $Z_{54}^-$\\
 & $-0.01150$& $0.9207$  &  $0.8680$ &  $-0.07719$ & $0.03252$ \\
\hline 
\hline
\end{tabular}
\end{center}
\end{table}

\clearpage
\newpage


\end{document}